\documentclass[floatfix,lengthcheck,showpacs,amssymb,amsmath,amsfonts,twocolumn,nofootinbib,longbibliography]{revtex4-1}
\usepackage{amsfonts,amsmath,units,wasysym,epsfig,graphicx,verbatim,color,subfigure,graphicx}
\usepackage{amsmath}
\usepackage{latexsym}
\usepackage{amssymb}
\usepackage{amsfonts}
\usepackage{mathtools}
\usepackage{bm}
\usepackage{color}
\usepackage{float}
\usepackage{tikz}
\usepackage{adjustbox}
\usepackage{array}
\usepackage{soul}
\usepackage{appendix}
\usepackage{physics}
\usepackage{braket}
\usepackage{xcolor}
\usepackage[none]{hyphenat}
\usepackage{academicons}
\usepackage{lineno}

\newcommand{\Real}{\operatorname{Re}}
\newcommand{\approximant}[1]{{\fontfamily{qcr}\selectfont{#1} }}

\def\be{\begin{equation}}
\def\ee{\end{equation}}
\def\bea{\begin{eqnarray}}
\def\eea{\end{eqnarray}}

\usepackage{academicons}
\definecolor{orcidlogocol}{HTML}{A6CE39}

\begin{document}
\author{\href{https://orcid.org/0000-0002-5077-8916}{Rahul Dhurkunde} and \href{https://orcid.org/0000-0002-1850-4587}{Alexander H. Nitz}}
\affiliation{Max-Planck-Institut fur Gravitationsphysik (Albert-Einstein-Institut), D-30167 Hannover, Germany}
\affiliation{Leibniz Universitat Hannover, D-30167 Hannover, Germany}

\date{\today}

\title{ Sensitivity of spin-aligned searches for neutron star-black hole systems using future detectors}

\begin{abstract}

Current searches for gravitational waves from compact-binary objects are primarily designed to detect  the dominant gravitational-wave mode and assume that the binary components have spins which are aligned with the orbital angular momentum. These choices lead to observational biases in the observed distribution of sources. Sources with significant spin-orbit precession or unequal-mass-ratios, which have non-negligible contributions from sub-dominant gravitational-wave modes, may be missed; in particular, this may significantly suppress or bias the observed neutron star -- black hole (NSBH) population. We simulate a fiducial population of NSBH mergers and determine the impact of using searches that only account for the dominant-mode and aligned spin. We compare the impact for the Advanced LIGO design, A+, LIGO Voyager, and Cosmic Explorer observatories. We find that for a fiducial population where the spin distribution is isotropic in orientation and uniform in magnitude, we will miss $\sim 25\%$ of sources with mass-ratio $q > 6$ and up to $\sim 60 \%$ of highly precessing sources $(\chi_p > 0.5)$, after accounting for the approximate increase in background. In practice, the true observational bias can be even larger due to strict signal-consistency tests applied in searches. The observation of low spin, unequal-mass-ratio sources by Advanced LIGO design and Advanced Virgo may in part be due to these selection effects. The development of a search sensitive to high mass-ratio, precessing sources may allow the detection of new binaries whose spin properties would provide key insights into the formation and astrophysics of compact objects.

\end{abstract}
    \maketitle

\section{Introduction}
    Since the first GW event in 2015, nearly 100 gravitational-wave (GW) observations have been made \cite{LIGOScientific:2021djp, Nitz:2021uxj, Olsen:2022pin}. The recent observing run (O3) of the Advanced LIGO  ~\cite{LIGOScientific:2014pky} and Virgo ~\citep{Acernese:2015gua} detectors primarily found binary black hole (BBH) mergers, in addition to a few notable events -- a new binary neutron-star (BNS) merger \cite{LIGOScientific:2020aai} and for the first time two neutron star-black hole (NSBH) mergers, GW200105 and GW200115 \cite{LIGOScientific:2021qlt}. NSBH systems could potentially produce an electromagnetic counterpart~\cite{Margalit:2017dij} and are expected to provide tight constraints on their spin components~\cite{LIGOScientific:2018jsj, Gompertz:2021xub}. The sensitivity of gravitational-wave observatories is rapidly increasing~\cite{KAGRA:2013rdx}; the Advanced LIGO detectors are expected to reach close to their design sensitivity in the next couple years \cite{LIGOScientific:2014pky, KAGRA:2013rdx} and eventually reach the A+ configuration  which will increase the detection rate of NSBH sources by $\sim 4-10 \times$\cite{Cahillane:2022pqm, LIGO:2020xsf}. Third generation detectors such as Cosmic Explorer \cite{Reitze:2019iox, Reitze:2019iox} are expected to be significantly more sensitive at low frequencies, and further improve the detection rate by several orders of magnitude~\cite{LIGO:2020xsf}.

    The origin of compact-binary mergers is still not well understood~\cite{Belczynski:2015tba,LIGOScientific:2016vpg}; the various formation channels can be broadly separated into dynamical or isolated scenarios \cite{LIGOScientific:2016vpg}. Binary sources with aligned spins and quasi-circular orbits are likely to have an origin within isolated environments \cite{Broekgaarden:2021hlu, Belczynski:2010tb, Dominik:2012kk}. Whereas, dense environments such as globular clusters often predict eccentric orbits or misaligned spins, which would cause spin-orbit precession \cite{Pooley:2003zb, Rodriguez:2015oxa}. A precessing system allows better measurement of the spin tilt-angles, and thus, carries an imprint of its evolutionary history \cite{Gompertz:2021xub, Stevenson:2017dlk, Talbot:2017yur, Rodriguez:2016vmx, Johnson-McDaniel:2021rvv}. Additionally, higher-order (HM) modes of the signal can help break mass-spin degeneracies \cite{Krishnendu:2021cyi}.

    In the recent compact-object merger catalogs \cite{LIGOScientific:2021djp, Olsen:2022pin, Nitz:2021uxj}, a few mergers exhibit precession or higher-order modes; in particular strong evidence of precession have been observed in the BBH merger GW200129~\cite{Hannam:2021pit}. Strong evidence for higher-order mode emissions have been observed for the events GW190814 and GW190412 ~\cite{LIGOScientific:2020stg, LIGOScientific:2020zkf}. The ring-down analysis of the event GW190521 indicates a subdominant mode \cite{Capano:2021etf}. 
    
    The current population of NSBH sources and precessing BBH sources challenge the various formation channels and population synthesis models \cite{Broekgaarden:2021efa}. Furthermore, the models have large uncertainties due to limited knowledge of the distribution of stellar metallicity \cite{Fryer:1999ht, Belczynski:2009xy} and natal kicks \cite{Lyne:1994az, Janka:2013hfa}. The detection of NSBH mergers could be decisive in constraining these models \cite{Broekgaarden:2021iew, Gupta:2017dsq}. However, current search methods are not optimised to capture precessing NSBH sources~\cite{Harry:2013tca, Usman:2015kfa, Messick:2016aqy, Aubin:2020goo, Hooper:2011rb}. 
    
    The most sensitive searches for gravitational-waves from compact-binary mergers use matched filtering \cite{Usman:2015kfa, Messick:2016aqy, Aubin:2020goo, Hooper:2011rb}, the core of which requires a bank of accurate signal templates. Several waveform models are available \cite{Pratten:2020ceb,Ossokine:2020kjp} which characterize the waveform in terms of 15 parameters, e.g. component masses, spins, sky location, etc.
    Naively searching over a 15 dimensional parameter space is computationally infeasible. 
    Typically, searches assume that the spins of the compact object are aligned with the orbital angular momentum, binary orbits are quasi-circular and only the dominant mode $(l, m) = (2,2)$ of the gravitational emission is observable.
    Because NSBH sources have high mass-ratios, it is expected that they can have non-negligible higher-order modes, and if their black hole is highly spinning, significant precession; searches that neglect these effects may have strong observational biases.
    
    Previous studies have examined the potential performance of dominant-mode only precessing searches~\cite{Harry:2016ijz} and the bias due to neglecting precession and higher-order modes for binary black hole searches~\cite{Harry:2017weg, Chandra:2022ixv, CalderonBustillo:2016rlt}. These searches require at least an order of magnitude more templates than typical analyses. Due to the increase in the number of templates, such analyses should be expected to produce a higher rate of false alarms at a fixed signal-to-noise (SNR) threshold. Despite the higher background, they observe a significant improvement in sensitivity at a fixed false alarm rate for sources with asymmetrical masses. In this paper, we determine the observational biases of aligned-spin searches to neutron star -- black hole mergers using an updated waveform model ~\cite{Pratten:2020ceb} that includes both the effects of precession and higher-order modes. We further study how this observational bias changes as gravitational-wave observatories improve.
    
    To study the detectability of wide range of NSBH sources, we simulate a population where $5 M_{\odot} \leq m_1^{det} \leq 30 M_{\odot}$ and $1 M_{\odot} \leq m_2^{det} \leq 3 M_{\odot}$ with an isotropic distribution of the spin angles and uniform distribution of spin magnitude. We compare the sensitivity of aligned spin searches in Advanced LIGO \cite{LIGOScientific:2014pky}, A+ \cite{Cahillane:2022pqm}, LIGO Voyager \cite{LIGO:2020xsf} and Cosmic Explorer \cite{Reitze:2019iox}. We determine the
    fraction of sources that would be detected by a dominant-mode, aligned-spin search compared to the ideal search that fully accounts for precession and higher-order modes. After approximating the increased background of the ideal search, we find dominant-mode, aligned-spin searches, such as those typically employed, will miss up to $\sim 25\%$ of NSBH systems with mass-ratios $q > 6$, and, up to $\sim 60\%$ of highly precessing $\chi_P > 0.5$ sources. This suggests, a significant observational bias against precessing sources whose detection can provide crucial information in understanding the formation channels and the astrophysics of compact-binary coalescing (CBC) sources.
    
    The paper is outlined as follows. In section \ref{sec:signal_model} we briefly introduce the signal model for CBC sources. We discuss our reference NSBH population in section \ref{sec:ref_pop}. In section \ref{sec:mock_searches} we describe the crux of modelled searches and metrics to define the search performance. In section \ref{sec:results}, we asses the loss in sensitivity for an aligned-spin search using the established metrics. In addition, we discuss briefly some challenges to develop a fully precessing search in section \ref{sec:challenges}. Finally, in sec VII we make concluding remarks.

\section{Modelling precessing CBC signals with HMs}\label{sec:signal_model}

Numerous models are available to describe the complete inspiral-merger-ringdown parts of the GW signal from compact-binaries~\cite{Pratten:2020ceb, Ossokine:2020kjp, Blackman:2017pcm}. These models can account for spin-precession effects and the higher-order modes of the signal, and are mainly categorized as -- phenomenological models \cite{Pratten:2020ceb}, effective one-body numerical relativity (EOBNR) models \cite{Ossokine:2020kjp} and surrogate models \cite{Blackman:2017pcm}. Currently, searches employ waveform models from both the phenomenological and effective-one body family ~\cite{LIGOScientific:2021djp, Nitz:2021uxj, Olsen:2022pin, Chandra:2022ixv}. Since our work primarily concerns the performance of model-based searches, we will briefly describe the signal model.

A quasi-circular CBC signal model is characterized by 15 parameters. The intrinsic parameters $\kappa$ of the system are the component masses $(m_1, m_2)$ and component spin vectors ($\vec{\chi}_1, \vec{\chi}_2$). The extrinsic parameters are the sky-location angles $(\alpha, \delta)$ in the frame of the observer, luminosity distance $d_L$, the inclination angle between the orbital angular momentum $\textbf{L}$ and the line of sight to the observer, polarization angle $(i, \psi)$, the time $t_c$ and the orbital phase $\varphi_c$.

The gravitational wave strain $h(t)$ as seen by a detector is a linear combination of the two gravitational-wave polarization 
\begin{align} 
    \begin{aligned}
    h(t) = F_{+}(\alpha, \delta, \psi) & h_{+}(\kappa, i, d_L, \varphi_c; t) \\ & + F_{\times}(\alpha, \delta, \psi)h_{\times}(\kappa, i, d_L, \varphi_c; t),
    \end{aligned}
    \label{Eq:Detector_response}
\end{align}
where the coefficients $F_{+}$ and $F_{\times}$ are the time-independent antenna pattern functions of the detector. The two polarization are defined in the radiation frame and together they make up the complex strain $H = h_{+} + i h_{\times}$. This complex strain can be further decomposed using the spin-2 weighted spherical harmonics $Y^{-2}_{lm}$
\begin{align}
        \label{Eq:spherical_harmonics}
    H \equiv h_+ + ih_{\times} = \sum_{l\geq2}\sum_{m=-l}^{m=l} Y^{-2}_{l,m}(i, \varphi_c) h_{l,m}(\kappa, d_L ;t-t_c),
\end{align}
where the $h_{l,m}$ are the various \textit{modes} of the GW signal. These various modes  (explicit expression can be found in \cite{Mills:2020thr})) have different contributions to the observed signal via the respective $Y_{l,m}(i, \varphi_c)$. The dominant mode $(l,m)$ = (2,2) is the strongest at face-on ($i=0$) or face-off ($i = \pi/2$) configurations and grows fainter with increasing $i$ reaching minimum at $i = \pi/2$. However, significantly inclined binaries can have stronger higher modes. In general for a given mode ($l, m$), we can write  
\begin{align}
    h_{l,m}(\kappa, d_L; t) = A_{l,m}(\kappa, d_L; t)e^{-i\Phi_{l,m}(\kappa;t)},
    \label{Eq:single_mode}
\end{align} 
where $A_{l,m}, \Phi_{l,m}$ are respectively the real amplitude and phase for a given mode. The phase $\Phi_{l, m}$ for each mode is related to the orbital phase as $\Phi_{l,m}(t) = m\phi_{orb}(t)$ up to a good approximation ~\cite{Apostolatos:1994mx}, which depends strongly on the component spins.

\begin{figure}[H]
    \centering
    \includegraphics[width=\linewidth]{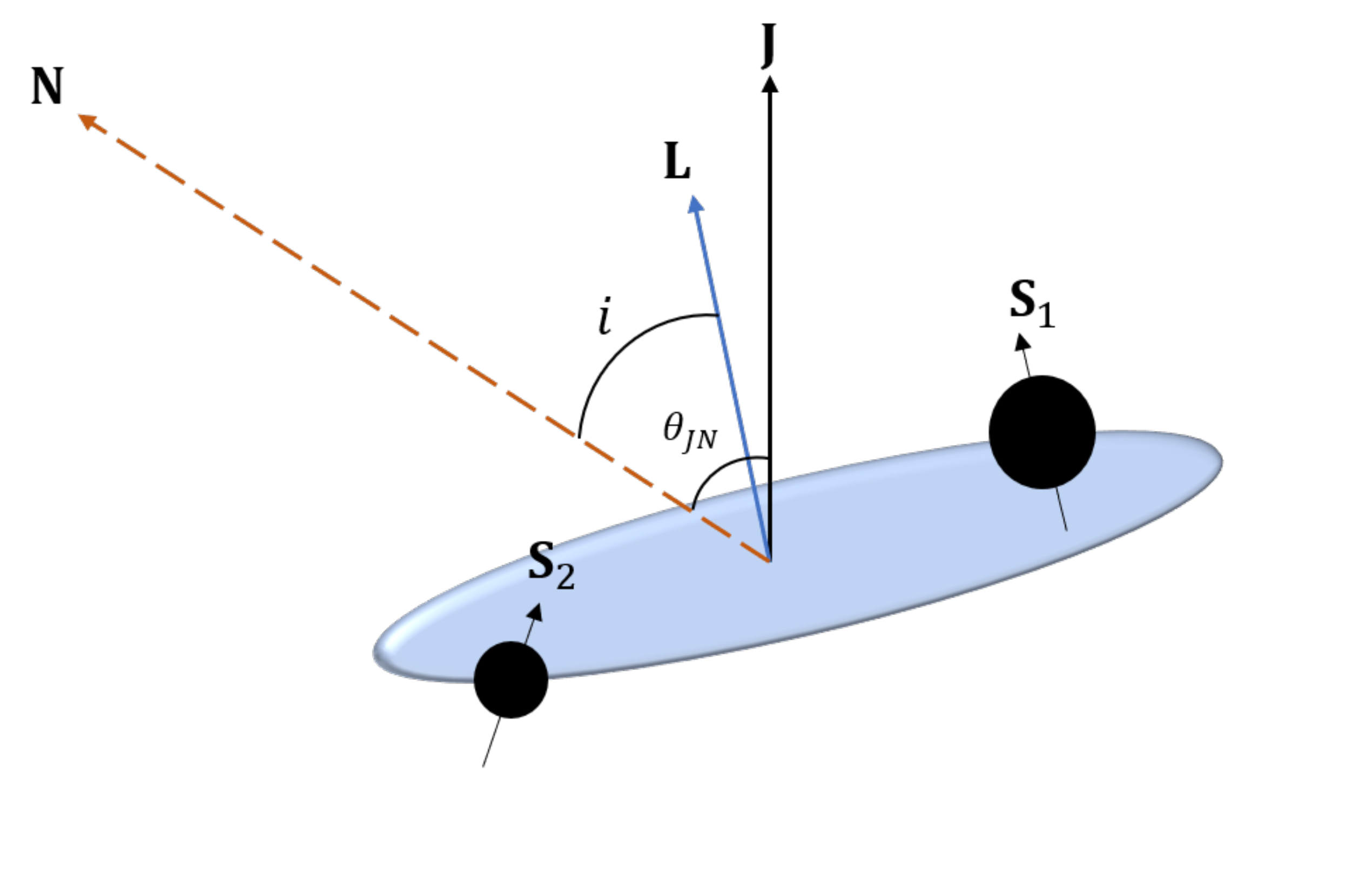}
    \caption{The angular momentum vectors for a precessing binary. The inclination angle $i$ is defined as the angle between the orbital angular momentum $\textbf{L}$ and the line of sight to the observer $\textbf{N}$. The angle between $\textbf{J}$ and $\textbf{N}$ is denoted as $\theta_{JN}$. Due to spin-orbit coupling, $\textbf{L}$ and $\textbf{S}$ will precess around the approximately fixed $\textbf{J}$.} 
    \label{fig:precessing_angles}
\end{figure}

The individual modes $h_{l,m}$ are functions of the spin parameters.
We denote the spin angular momenta $\textbf{S}_1 = \vec{\chi}_1 m_1^2$ and $\textbf{S}_2= \vec{\chi}_2 m_2^2$.  In the case of aligned-spin systems, the direction of the orbital angular momentum remains fixed and thus, $\textbf{S}_1 \textbf{S}_2$, and $\textbf{L}$ are parallel. In the case of systems with generically oriented spin vectors, the spins of the compact objects couple with the orbital angular momentum which may cause spin-precession. For such systems, the orbital angular momentum $\textbf{L}$ precesses around the nearly fixed total angular momentum $\textbf{J} = \textbf{S}_1 + \textbf{S}_2 + \textbf{L}$; the inclination angle varies over time. The various spin and orientation angles are shown in the Fig. \ref{fig:precessing_angles}. Note, even though $\textbf{J}$ is considered to be fixed to a good approximation \cite{Apostolatos:1994mx}, there can be rare instances where $\textbf{J}$ can change significantly \cite{Apostolatos:1994mx}. 

Precession dynamics cause phase and amplitude modulations to the observed signal. Since precession is caused by the non-alignment of the spins with the orbital angular momentum, to measure the strength of precession, it is common to group the in-plane and parallel spin components. The spin effects are commonly characterized in terms of the two effective parameters \cite{Ajith:2009bn, Schmidt:2014iyl}

\begin{align}
    \chi_{\text{eff}} = \dfrac{1}{M}\Bigg(\dfrac{\textbf{S}_1}{m_1}+\dfrac{\textbf{S}_2}{m_2}\Bigg)\cdot\hat{\textbf{L}},\\
    \chi_{p} = \dfrac{1}{A_1m_1^2}\max\Big(A_1 S_1^{\perp},A_2 S_2^{\perp}\Big),
\end{align} 

where, $A_1=2+3q/2$ and $A_2 = 2 + 3/(2q)$, $q$ is the mass-ratio $m_1/m_2$, and $S_i^{\perp}$ is the projection of the spins orthogonal to $\textbf{L}$. The $\chi_{\text{eff}}$ parameter gives us a measure of the spin components parallel to $\textbf{L}$. The four in-plane spin components are averaged over a precessing cycle to obtain an effective $\chi_p$ precession parameter.

\section{ Reference NSBH population } \label{sec:ref_pop}
Since the precession and higher mode effects grow with mass-ratio, we choose to study a population of NSBH source extending up to $q = 20$ with the broad priors shown in table~\ref{tab:priors}. We sample the component masses in terms of the detector frame masses $m_{1,2}^{det} = (1+z)m_{1,2}$, where $z$ is the redshift of a source. The redshifted masses are sampled uniformly between $5 M_{\odot} \leq m_1^{det} \leq 30 M_{\odot}$ and $1 M_{\odot} \leq m_2^{det} \leq 3 M_{\odot}$. Working in the detector frame eliminates the redshift dependence in the calculation of signal-to-noise ratio and related metrics (see e.g. sec \ref{sec:metrics}).

We choose the spin directions to be isotropically distributed and the spin magnitudes to be uniform. We assume neutron-stars are slowly spinning with spin-magnitude up to 0.05, and up to one for black-holes. We also assume the sources are isotropically distributed in the sky. The polarization angle, coalescence phase and the cosine of the inclination angle are also uniformly distributed. In total we simulate a population of 50,000 compact-binaries.

In this study, we employ one of the latest models from the phenomenological waveform family \approximant{IMRPhenomXPHM} \cite{Pratten:2020ceb} to generate signals for our fiducial population. This model accounts for generic spins and HMs and includes $(l, |m|) = (2, 2), (2, 1), (3, 3), (3, 2), (4, 4)$ modes. \approximant{IMRPhenomXPHM} has shown consistent results with other waveform models in the recent compact-merger catalogs \cite{LIGOScientific:2021djp, Nitz:2021uxj, Olsen:2022pin}, inference studies of various GW events \cite{Estelles:2021jnz, Krishnendu:2021cyi}, and studies inferring population properties \cite{Tiwari:2021yvr, Zhu:2021jbw}. Thus, the reliability and computational efficiency of \approximant{IMRPhenomXPHM} motivated us to employ this model.  

\begin{table}[]
    \centering
    \begin{tabular}{c|c}
        Parameter &  Distribution\\
        \hline
        \hline
        $ m_{1}^{det} $ & uniform $\in$ [5.0, 30.0]  $ M_{\odot}$ \\
        \hline
        $ m_{2}^{det} $ & uniform $\in$ [1.0, 3.0] $ M_{\odot} $\\
        \hline 
        $\abs{\chi_{1}}$ & uniform $\in$ [0, 1] \\
        \hline
        $\abs{\chi_{2}}$ & uniform $\in$ [0, 0.05] \\
        \hline
        Spin tilt angles & Isotropic\\
        \hline 
        Sky angles & Isotropic \\
        \hline
        $\varphi_c$ & uniform $\in$ [0, $2\pi$]\\
        \hline
        $\cos{i}$ & uniform $\in$ [-1, 1] \\
        \hline
        $\psi$ & uniform $\in$ [0, $2\pi$]
    \end{tabular}
    \caption{ Various source parameters (first column) used in the simulation for the reference NSBH population consisting of 50,000 compact-binaries. The second column shows the distribution used to sample the corresponding parameter.}
    \label{tab:priors}
\end{table}

\section{Modelled Gravitational-wave Searches}\label{sec:mock_searches}

Searching for CBC signals is done in multiple stages which begins with filtering inteferometric data with a bank of templates to identify possible candidates ~\cite{Usman:2015kfa, Nitz:2018imz}. Then candidate events are followed up with signal consistency tests \cite{LIGOScientific:2017tza, Allen:2004gu}, data quality checks \cite{LIGOScientific:2017tza, LIGOScientific:2016gtq, LIGO:2021ppb}, and are finally assigned a statistical significance value \cite{Nitz:2018imz, LIGOScientific:2018mvr}. Modelled searches for CBCs use matched filtering and signal consistency tests which are sensitive to the employed waveform model. 

Current searches for compact-binaries make assumptions to the signal model which physically restrict the systems to have no eccentricity, no precession, and no observable HMs. As a consequence, these searches have a bias for aligned-spin sources and for sources which with only small higher-mode contributions, those that are nearly face-on/off or close to equal-mass. In this section we discuss the essentials of how an aligned-spin modelled gravitational-wave search is conducted. We introduce the detection statistic used by modelled searches, the different detector configurations used in this study, and briefly describe the method for obtaining an aligned-spin template bank containing only the dominant mode, and finally the metrics used to assess the performance of an aligned spin search using a specific template bank and assuming a particular detector sensitivity curve.

\subsection{Search statistic}

Modeled searches for GWs employ \textit{matched filtering} to the detector data. The matched filter is an optimal statistic to detect an anticipated signal in stationary Gaussian noise \cite{Allen:2005fk}. It is computed in the Fourier domain by correlating the data $\tilde{s}(f)$ with the signal model $\tilde{h}(f)$ weighted by the noise power spectral density (PSD) $S_n(f)$; the complex matched filter statistic is 
\begin{align}
    \label{Eq:matched_filter}
    \braket{s|h} = 4\int_{0}^{\infty} \dfrac{\tilde{s}(f)\tilde{h}^*(f)}{S_n(f)}.
\end{align}
The signal-to-noise (SNR) ratio $\rho$ which is the matched-filter output maximised over an overall amplitude,   
\begin{align}
    \label{Eq:SNR_def}
    \rho^2 = \dfrac{(\Real[\braket{s|h}])^2}{\braket{h|h}}.
\end{align}

Since the signal parameters are unknown, the SNR is maximised over the parameter space. A naive maximisation procedure over the complete 15-dimensional parameter space is computationally challenging. Therefore, the component spins are typically assumed to be aligned with the orbital angular momentum and a search is conducted only for the dominant gravitational-wave mode. Under these assumptions the two polarizations of the signal are related by a simple phase shift $\tilde{h}_+(f) = i \tilde{h}_{\times}(f)$. Using Eqs.(\ref{Eq:Detector_response}) and (\ref{Eq:spherical_harmonics}) the signal seen by the detector in Fourier domain is simplified to 
\begin{align}
    \label{Eq:maxmising_templates}
    \tilde{h}(f) = A(f)e^{i \phi_0}\tilde{h}_0(\kappa; f)e^{2\pi i f t_c},
\end{align}
where $\tilde{h}_0(\kappa)$ depends only on the intrinsic parameters and the extrinsic parameters are factored out as the nuisance parameters -- an overall amplitude $A$ and phase $\phi_0$. 

As per Eq. (\ref{Eq:maxmising_templates}) the SNR is maximised in three categories of parameters -- 1) Intrisinc parameters $\kappa$, 2) fiducial parameters $A$ and $\phi_0$ 3) time of arrival $t_c$. The intrinsic parameters are searched by using a set of discrete points laid out on the four dimensional parameter space $\kappa^{||} = (m_1, m_2, \chi_{1z}, \chi_{2z})$. Waveforms evaluated with parameter values at a given sampled point is referred as templates and together they make up a template bank. The matched filter is repeatedly computed over all the templates to find the best matching template with highest SNR. Simultaneously, for each template, the SNR is maximised over the extrinsic parameters $(D, i, \psi, \alpha, \delta, \varphi_c)$ via the nuisance parameters ($A, \phi_0$) -- first by normalizing the matched filter with the power of the signal $\braket{h|h}$, and then using a quadrature of the SNR to maximise the unknown phase $\phi_0$. The maximization over these nuisance parameters is written as 
\begin{align}
    \mathop{\max}_{\phi_{0}}(\rho^2) = \dfrac{1}{2}\norm{\braket{s|\hat{h}_0}}^2,
    \label{Eq:simplified_statistics}
\end{align}
where $\hat{h}_0 = \tilde{h}_0/\braket{\tilde{h}_0|\tilde{h}_0}^{1/2}$. Finally, the position of the signal is efficiently searched over by performing an inverse fast Fourier transformation (FFT) to obtain the SNR time-series.

\subsection{Aligned-spin template banks for different detectors}
Constructing a template bank involves sampling discrete points in the parameter space such that any random point chosen in the allowed region must have atleast one sampled point within a fiducial distance $d_{max}$ which corresponds to the \textit{mismatch} between the waveforms computed at those two points. The value of the mismatch controls the overall density of the template bank which needs to be tuned carefully -- a high value leads to fewer templates and loss of signal SNR, on the other hand, a low value leads to larger number of templates which increases the computational cost of matched filtering. Hence, the template placement problem is to minimize the number of templates while maintaining a fixed minimum match (1-mismatch). There are numerous methods to generate a template bank which are essentially categorized into three categories -- geometric lattice based methods \cite{Owen:1998dk, Babak:2006ty}, stochastic placing algorithms \cite{Harry:2009ea, Babak:2008rb}, and hybrid methods \cite{Roy:2017oul, Capano:2016dsf}. We employ the stochastic sampling method to generate aligned-spin template banks.

\begin{figure}[H]
    \centering
    \includegraphics[width=\linewidth]{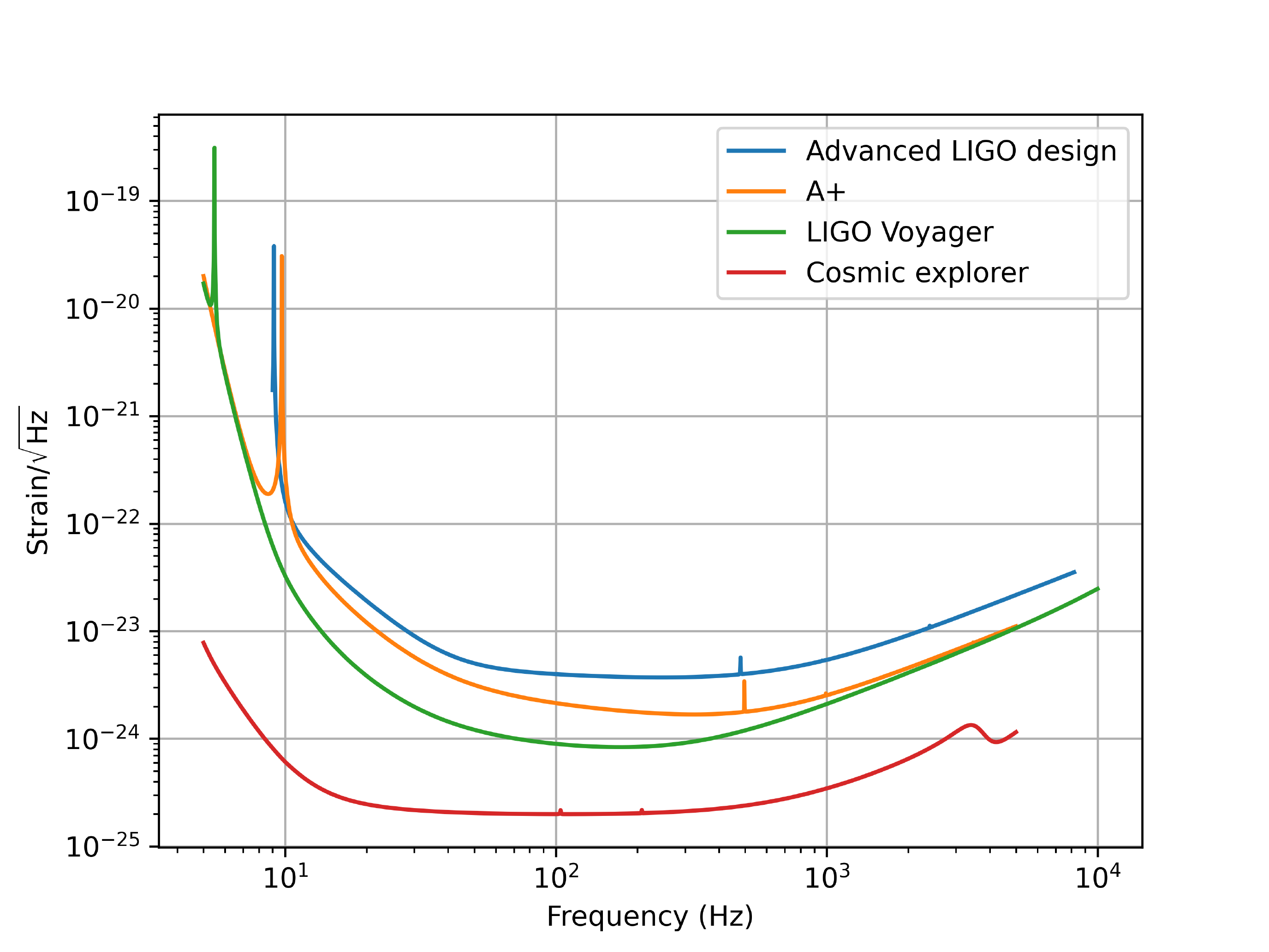}
    \caption{A comparison of the noise amplitude spectral density (ASD) of each detector configuration used in this study. Aligned-spin template banks are generated for each noise curve using the appropriate $f_{low}$. } 
    \label{fig:PSDs}
\end{figure}

The PSD (noise curve) of a detector can change the template bank by influencing the distribution of the sampled points or the size of the bank. For future detectors configurations, the sensitive bandwidth is expected to increase (see Fig.~\ref{fig:PSDs}). With wider bandwidths, a template can accrue greater phase mismatch with a slight variation in the parameters and this will result in increased number of templates. Both Advanced LIGO and A+ observatories are expected to be sensitive down to 15 Hz. LIGO Voyager is anticipated to be sensitive from 10 Hz onwards. Cosmic Explorer is predicted to be sensitive down to 5Hz; however, in \cite{Lenon:2021zac} the authors find $99.53\%$ of the signal SNR will be retained at lower frequency cutoff = 7Hz and Doppler modulations can be neglected from this lower frequency.  

In this work, for each detector configuration we generate only dominant mode, aligned-spin banks by sampling the 4D $(m_1, m_2, \chi_{1z}, \chi_{2z})$ parameter space using a stochastic sampling method. We use \approximant{SEOBNRv4\_ROM} model and use the appropriate lower frequency cutoff $f_{low}$ for a given sensitivity curve. All the template banks are generated with an average mismatch of $< 1\%$.  

\subsection{Metrics to quantify the detectability}\label{sec:metrics}
Searches with a template bank will inevitably lose a fraction of the population due to two main reasons -- discreteness of the template bank, and the templates not being the optimal description of signals. The loss in sensitivity is quantified using metrics that measure the bank's ability to recover a population of simulated signals. We first define the match between a given template $h_i(\Phi)$ with unknown extrinsic parameters $\Phi$ and a signal $g$ in terms of the overlap $\braket{g|h_i({\Phi})}$ between the two as
\begin{align}
    m(g, h_i) \equiv \mathop{\max}_{\Phi}\Big( \braket{g|h_i(\Phi)} \Big) = \dfrac{1}{\sqrt{2}}\norm{\braket{g|h_i}},
    \label{eq:overlap}
\end{align}
where $i$ is the index corresponding to different templates in a given bank. Computing the match involves maximization over all extrinsic parameters such as $\Phi = (D, i, \psi, \alpha, \delta, \varphi_c)$ that are not included in the template bank. Since the intrinsic parameters remain unknown, the match is further maximised over the template bank to obtain the best-fit template to $g$ and its \textit{fitting factor} 
\begin{align}
    \text{FF}(g) = \max_{i} \big(m(g,h_i)\big).
    \label{Eq:FF}
\end{align}

The fitting factor corresponds to the maximum fraction of the absolute signal SNR that can be recovered by a given template bank which ranges $\in [0, 1]$. For instance, a signal with an SNR = 10.0 having a FF of 0.98 can only be detected with a maximum SNR of 9.8. 

The number of signals detected by a search above some SNR threshold, depends not only on the FF achieved but also on the intrinsic loudness $\sigma(s_i)$ of each signal $s_i$. Often, systems with poor FFs are also intrinsically quieter. We use a metric (introduced in \cite{Buonanno:2002fy}) that takes into account the signal SNR and estimates the number of signal recovered relative to an optimal search that can perfectly match each signal. For a volumetric distribution of $n_s$ sources, the number of signals recovered by a search with a template bank $B$ of fitting factors ($\text{FF}_B$) is proportional to $\sum \limits_{i=0}^{n_s-1} \text{FF}_B^3(s_i)\sigma^3(s_i)$. The fraction of signals recovered relative to an optimal search with $\text{FF} = 1$ is referred as the signal recovery fraction (SRF) and is given by
\begin{align} 
    \text{SRF} \equiv \alpha^B = \dfrac{\sum \limits_{i=0}^{n_s-1} \text{FF}^3_B(s_i)\sigma^3(s_i)}{\sum \limits_{i=0}^{n_s-1} \sigma^3(s_i)}.
    \label{eq:SRF}
\end{align}

For a given search, the false-alarm-rate (FAR) can be uniquely mapped to SNR threshold, and detections are often determined by their statistical significance exceeding a false alarm rate threshold. This mapping can vary with different searches as the background distribution of triggers changes. The relative sensitivity between two searches ensuring the same significance is compared at a fixed FAR. Using the threshold $\rho_0$ corresponding to a chosen FAR value, the SRF weighted by the FAR can then be computed using $\alpha^{B}/\rho_{0}^3$. As an example, consider two different searches scenarios -- $B$ (dominant mode, aligned-spin) and $B'$ (including precession or HMs) with SNR thresholds $\rho_{\text{ref}}$ and $\rho_{\text{thresh}}$ respectively at a fixed FAR. The relative sensitivity between the two searches is computed according to
\begin{align}
    \beta = \dfrac{\alpha^B\rho^3_{\text{thresh}}}{\alpha^{B'}\rho^3_{\text{ref}}}
    \label{eq:relative_sensitivity}.
\end{align}

\section{Assessing the loss due to omitting HM and precessional effects}\label{sec:results}

In this section we estimate the loss in sensitivity for a dominant-mode, aligned-spin search to our reference NSBH population. We sample the source parameters according to table \ref{tab:priors} and generate precessing signals with HMs. To study the effects of precession and HMs in isolation, we create
two additional sets of signals -- precessing signals with only the dominant mode, and nonprecessing signals with HMs by setting the in-plane spin components to zero which ensures consistent $\chi_{eff}$ distribution. To validate the expected recovery of signals similar to the dominant mode aligned spin templates, we also generate nonprecessing signals with only the dominant mode. In summary, we create four different classes of signals -- nonprecessing dominant mode only (baseline), nonprecessing with HMs,  precessing dominant mode only, and precessing with HMs. We use \approximant{IMRPhenomD} \cite{Husa:2015iqa, Khan:2015jqa} to generate the baseline signals and \approximant{IMRPhenomXPHM} for all other classes of signals. All signals are projected to the detector frame by linearly combining the two polarizations with the antenna pattern functions as per Eq. (\ref{Eq:Detector_response}). 

\begin{table}[H]
    \centering
    \begin{tabular}{c|c|c}
         Class & Type & Waveform model \\
              &        &    \\
        \hline 
        \hline
       I & nonprecessing dominant mode only& \approximant{IMRPhenomD} \\
       \hline
       II & nonprecessing with HMs & \approximant{IMRPhenomXPHM} \\
       \hline
       III & precessing dominant mode only & \approximant{IMRPhenomXPHM} \\
       \hline
       IV & precessing with HMs & \approximant{IMRPhenomXPHM} \\
    \end{tabular}
    \caption{Four different classes of signals used in this work to study the impact of the missing features. The baseline signals (type I) are simulated using a nonprecessing waveform model without the HMs (\approximant{IMRPhenomD}), while all the other types are generated using an updated model which accurately models precession and HM effects (\approximant{IMRPhenomXPHM}) \cite{Pratten:2020ceb}.}
    \label{tab:injection_classes}
\end{table}

\begin{figure}[H]
    \centering
    \includegraphics[width=\linewidth]{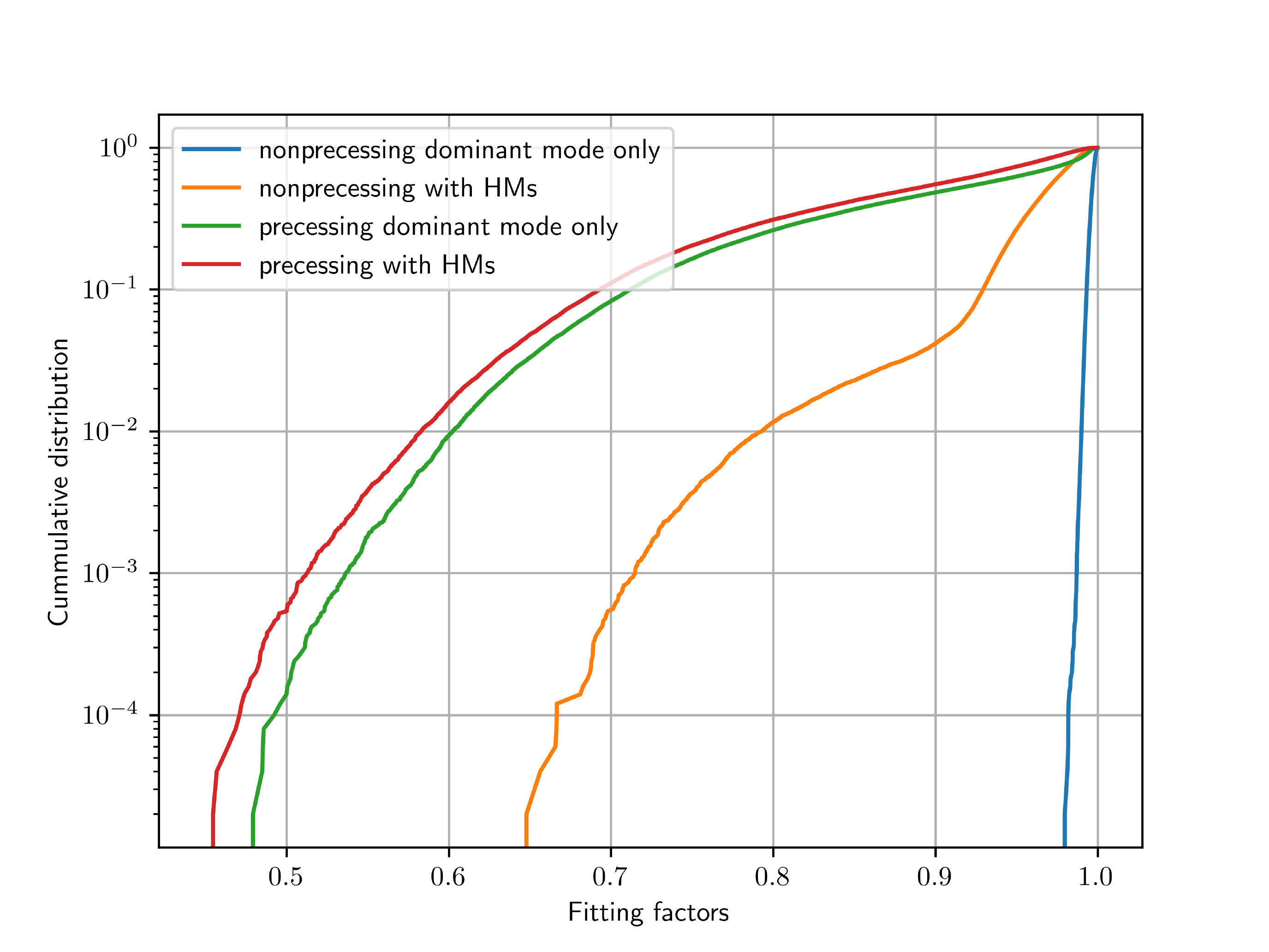}
    \caption{Cumulative distribution of the FFs for different classes of signals obtained using a dominant mode, aligned-spin template bank using the Advanced LIGO design noise curve. The blue curve confirms the expected number of signals recovered with high FFs above the average minimum match ($< 1\%$). We find long tails of poor FFs for signals with precession or HMs. Also, note precession effects dominate the FFs.}
    \label{fig:design_CDF}
\end{figure}

\subsection{signal-to-noise recovery}
 We calculate the distribution of fitting factors (fraction of SNR recovered by a reference aligned-spin search) for each class of signals. The cumulative distributions of the FFs for all classes of signals is shown in Fig. \ref{fig:design_CDF}. Since the templates and baseline injections belong to the same class of signals, as expected, we find the average FF is close to unity. However, the dominant mode, aligned-spin template bank shows long tails of poor FFs for signals with precession or HM effects; FF values down to $\sim0.5$ for precessing signals and down to $\sim0.65$ for nonprecessing signals with HMs. Higher-modes alone do not reduce the FF distribution as much as precession alone (without higher-modes). We expect the distribution of FFs to vary as a function of component mass, spin, and binary orientation, due to the non-uniform impact of precession and higher-order modes; we will study this variation in the following sections.

\subsubsection{Component Masses}
\begin{figure*}[]
    \centering
    \includegraphics[width=13cm, height=10cm]{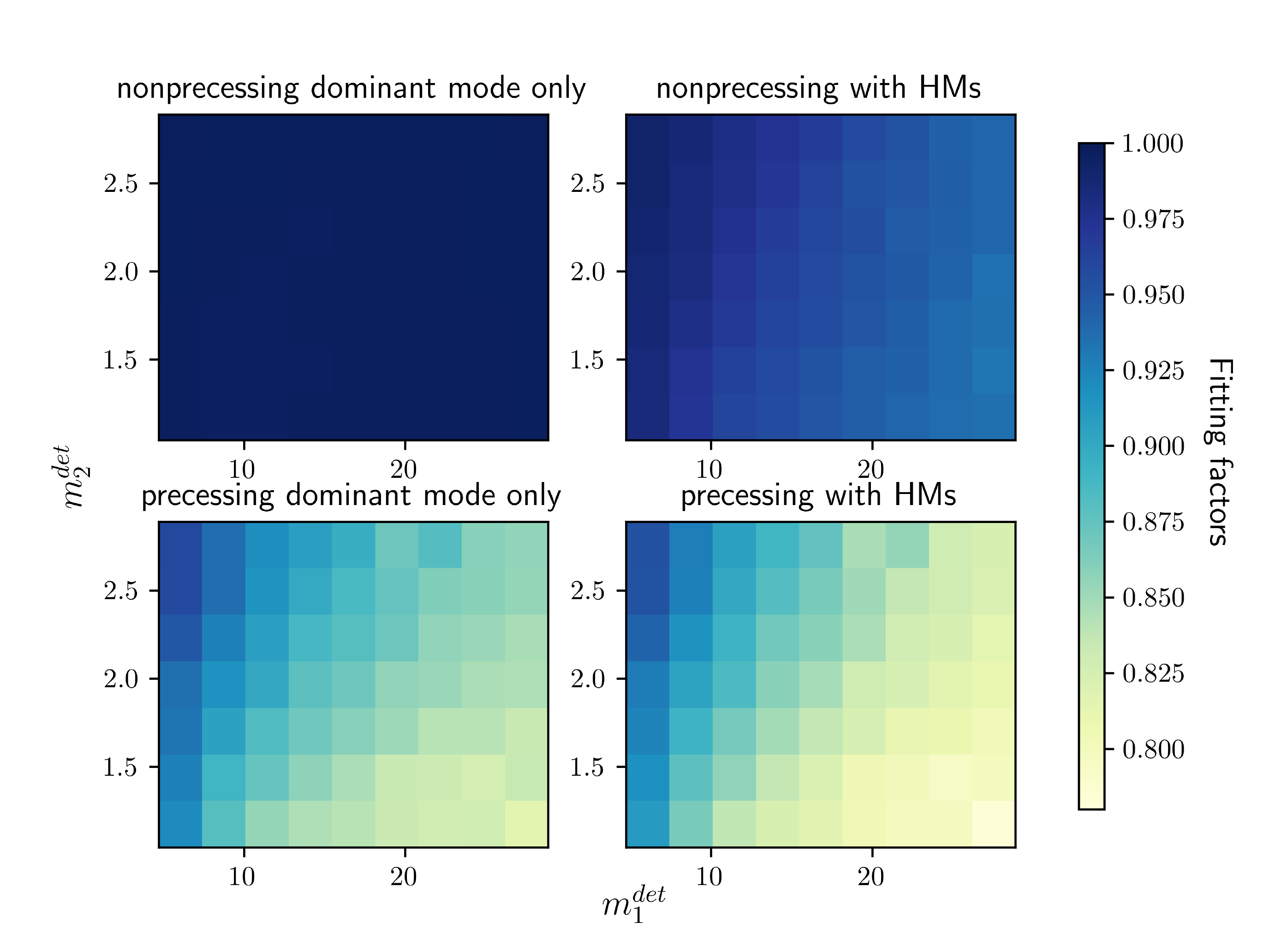}
    \caption{Average fitting factors for a population of simulated NSBH signals distributed across the ($m_1^{det}, m_2^{det}$) space obtained using a dominant mode, aligned-spin template bank and assuming the Advanced LIGO design sensitivity. We show the FFs for four different class of signals -- nonprecessing dominant mode only(top-left), nonprecessing with HMs (top-right),  precessing dominant mode only (bottom-left) and  precessing with HMs (bottom-right). We observe a reduction in FFs (fractional loss in observable SNR) up to $8\%$ for signals with HMs, and up to $20 \%$ for precessing sources. Notice the precession effects dominate the FF distribution over HMs.}
    \label{fig:FF_m1m2}
\end{figure*}

 The impact of both precession and HMs on the observed gravitational-wave signal grows with the mass-ratio of the system. In Fig. \ref{fig:FF_m1m2} we show the population-averaged FF for all classes of signals for the Advanced LIGO design sensitivity as a function of $(m_1^{det}, m_2^{det})$. The top left subplot confirms the intended recovery of baseline signals throughout the parameter space. For nonprecessing signals with HMs (top right), we notice only a slight drop in the FF values (up to $\sim$ 8\%). With precessing signals (bottom row) we observe a clear trend in the FF distribution; FFs decrease with increasing mass-ratio (top-left to bottom-right). We estimate the average fractional loss in SNR can be as high as $\sim 15 - 22\%$ for highly asymmetric precessing systems. By comparing the subplots in the bottom row from Fig. \ref{fig:FF_m1m2}, we infer the precession effects dominate the FF distribution. These results confirm that HMs and precession effects grow stronger with increasing mass-ratios of compact-binaries.

\subsubsection{Orientation of the binary}

\begin{figure}[h]
    \centering
    \includegraphics[width=\linewidth, height=5.8cm]{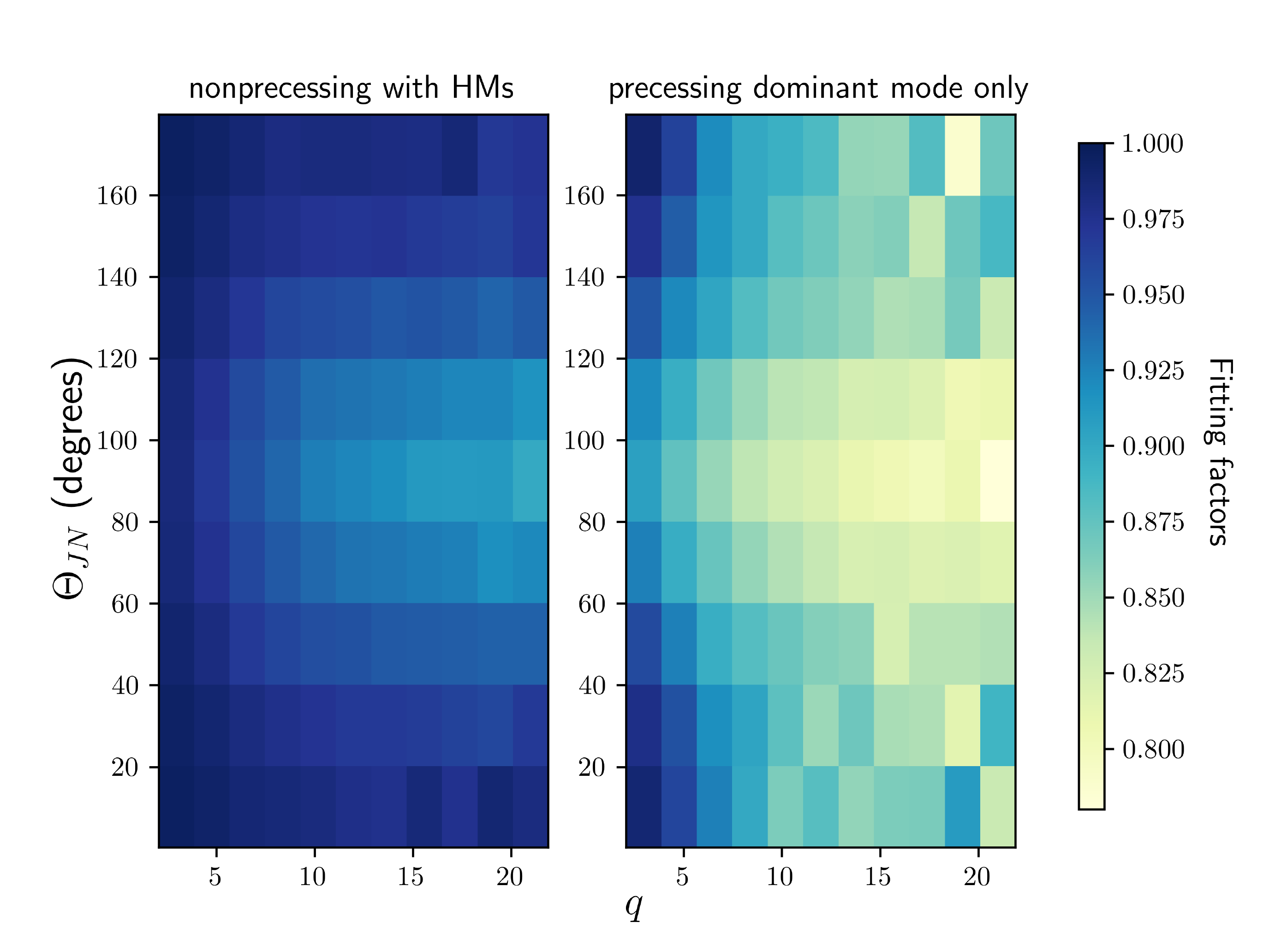}
    \caption{ The average FFs as a function of ($q, \theta_{JN}$) at $f_{ref} = 100$ Hz, for a simulated population of signals containing HMs  (left) and precession with only the dominant mode (right). We observe larger losses in FFs for precessing signals than for signals with HMs. We notice a trend of decreasing FFs when the source orientation changes from face-on/off to edge-on or when the mass-ratio increases.}
    \label{fig:FF_orientation}
\end{figure}

The orientation of a binary is generally characterized in terms of the inclination angle $i$ between $\textbf{L}$ and $\textbf{N}$. However, in the case of precessing system $i$ changes over time. To study the dependence of FFs on the orientation of the binary, we instead estimate the distribution in terms of $\theta_{JN}$ at a reference frequency $f_{ref}$ = 100 Hz; $\theta_{JN}$ will typically be stable over the duration of the signal. In Fig. \ref{fig:FF_orientation} we show the binned population-averaged FF distribution as a function of $(q,\theta_{JN})$ for nonprecessing signals with HMs (left) and dominant mode precessing signals (right). Similar to Fig. \ref{fig:FF_m1m2}, we confirm both HMs and precession effects grow stronger with increasing mass-ratios. Importantly, FFs decreases as the binary moves away from face-on/off configuration and reaches minimum at the edge-on configuration. This is expected because the dominant mode is faintest at $i=\pi/2$ and also various higher-order modes have significant contribution for edge-on binaries. 
For the non-precessing, dominant-mode only sources, a maximum loss in FFs up to $10\%$ for highly asymmetric nearly edge-on binaries. In the right plot, we observe much larger losses in FFs for precessing systems. We observe minor losses up to $\sim 10\%$ even for nearly equal-mass systems. For precessing systems with large mass-ratios ($q > 10$), we observe losses in FFs up to $\sim 18-23\%$ for edge-on and up to $\sim 16\%$ for face-on/off binaries.

\subsubsection{Component spins}
\begin{figure}[]
    \centering
    \includegraphics[width=\linewidth, height=5.8cm]{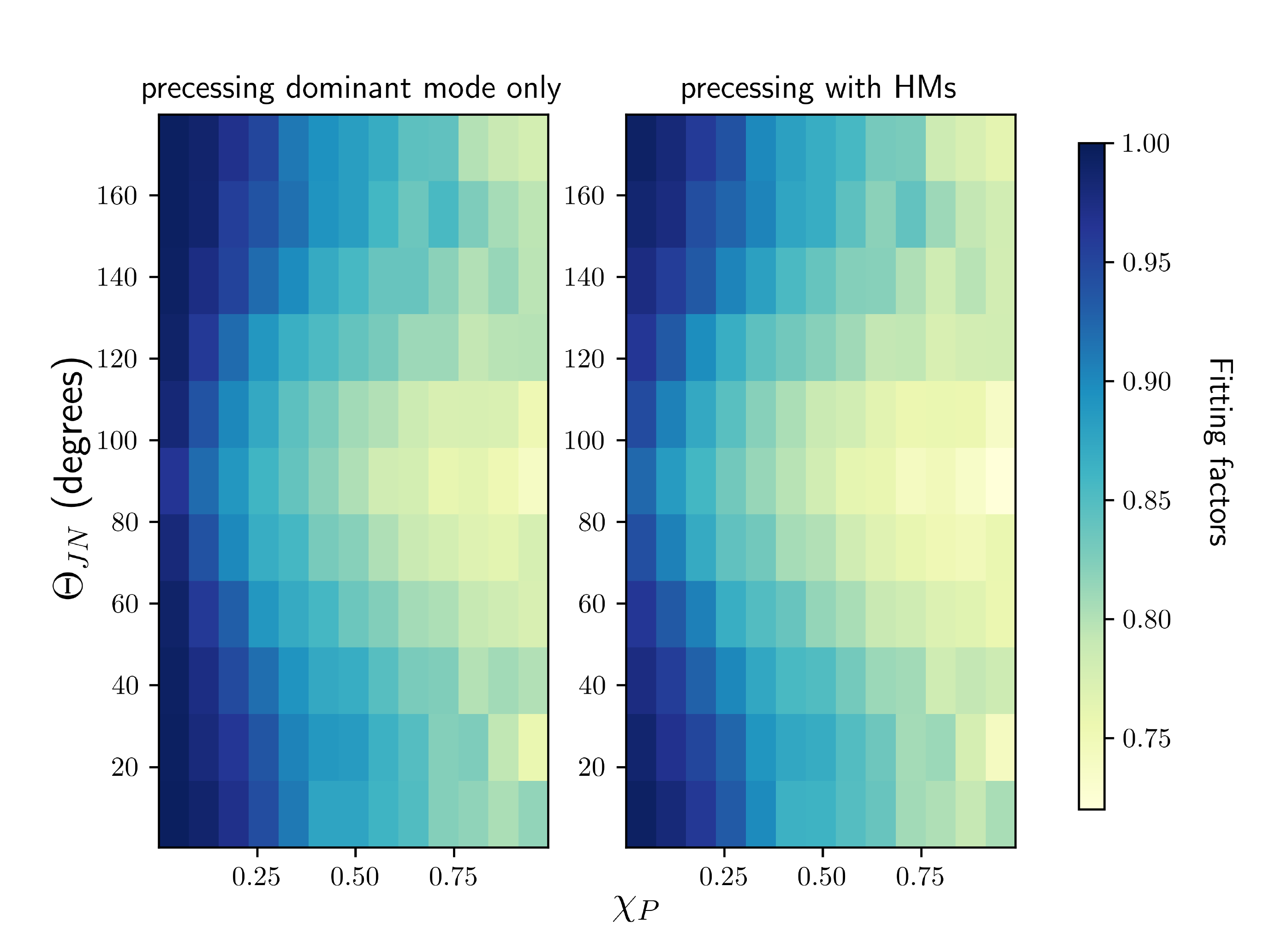}
    \caption{ The average FFs as a function of effective precession and source orientation $(\chi_p, \theta_{JN})$, for precessing signals without HMs (left) and with HMs (right). Due to the dominance of the precessing effects, there is no significant difference between the two subplots. The  FFs decrease as the effective precession increases and the source becomes closer to edge-on, up to an average loss of $\sim20-30\%$ in SNR.}
    \label{fig:chi-theta}
\end{figure}

 The ability for in-plane spin to cause precession, is typically measured in terms of the effective precession $\chi_P$ parameter \cite{Schmidt:2014iyl}. The binned population-averaged FF distribution as a function of $(\chi_P, \theta_{JN})$ is shown in Fig. \ref{fig:chi-theta}. We use two classes of signals, precessing signals without HMs (left), and with HMs (right). As expected from Fig. \ref{fig:FF_orientation}, for a fixed value of $\chi_P$ we observe the same trend across $\theta_{JN}$ -- decreasing FFs as the binary shifts away from face-on or face-away configuration. As the precession increases (left to right), the FFs decrease; this occurs because precession causes strong amplitude and phase modulations in the signal which which may not be captured by nonprecessing templates. The nearly identical subplots suggest insignificant impact from the HMs. For slightly precessing sources ($\chi_P < 0.5$), we find $\sim 12-20\%$ loss in FFs. For highly precessing systems the loss in FFs can be up to $\sim 28\%$ for edge-on and up to $\sim 20\%$ for face-on binaries.

\subsubsection{Comparing different detector configurations}

\begin{figure*}[]
        \centering
        \includegraphics[width=13cm, height=10cm]{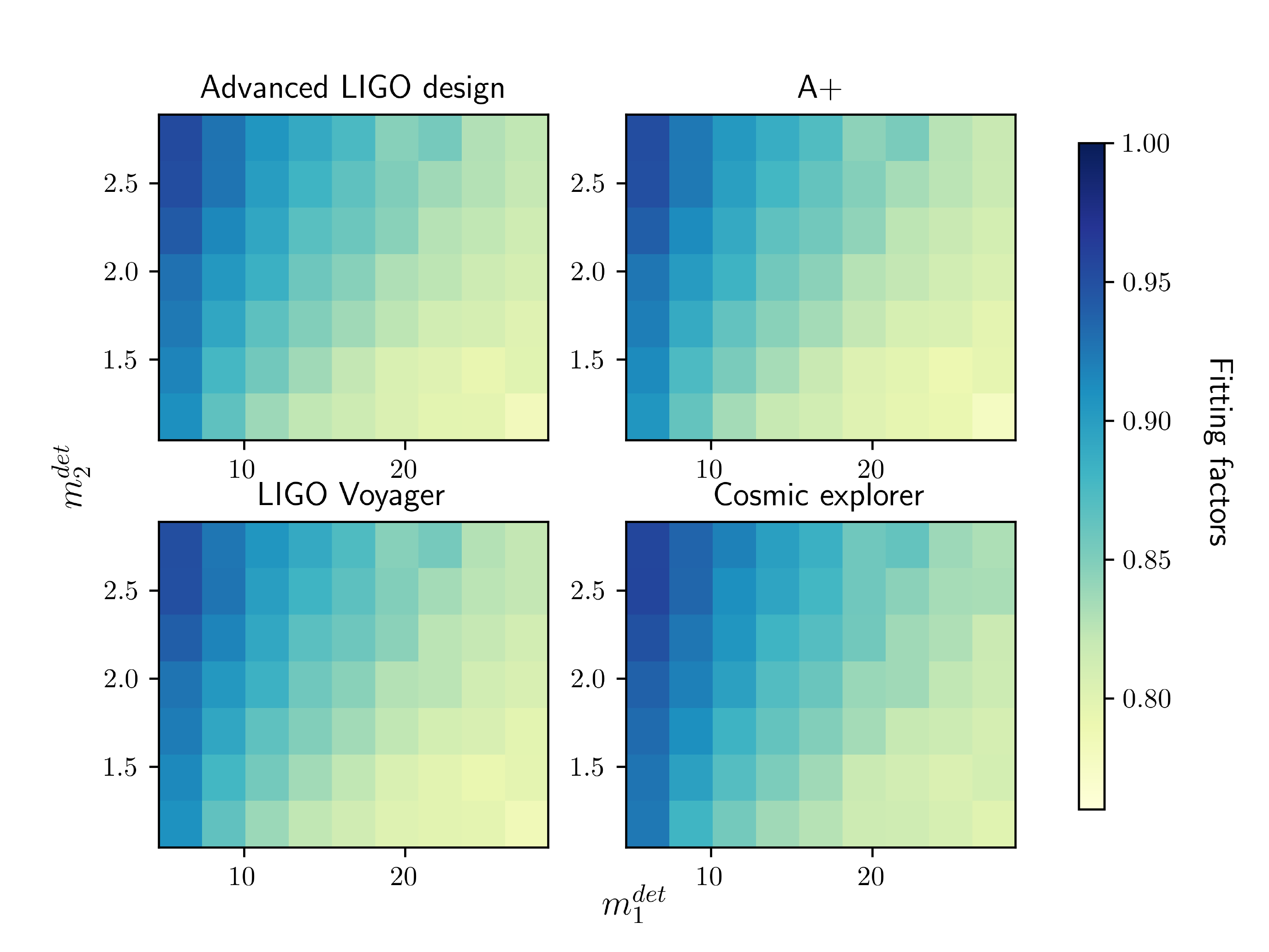}
        \caption{ FF distribution for precessing sources with HMs as a function of $(m_1^{det}, m_2^{det})$ for different detector configurations. We can see that the results are not strongly dependent on the noise curve.}
        \label{fig:FF_diffPSDs.png}
\end{figure*}

Fitting factor depends on the noise curve (PSD) and the lower frequency $f_{low}$ used to compute matches as per equations (\ref{eq:overlap}) and (\ref{Eq:FF}). We study the variation in FF distribution across four different detector configurations by using their respective noise curves (as shown in Fig. \ref{fig:PSDs}) and the appropriate $f_{low}$. 
In the Fig. \ref{fig:FF_diffPSDs.png}, we show the distribution of the FFs across ($m_1^{det}, m_2^{det}$) space for all detector configurations. To aid in comparison, we show the FFs for only the precessing signals with HMs case. There is no significant difference between FFs for the different detectors and detector configurations. The change in the noise curve between Advanced LIGO though Cosmic Explorer is not large enough to significantly change the key results.

\subsection{Number of Detectable Sources}

\begin{figure*}[]
        \centering
        \includegraphics[width=13cm, height = 10cm]{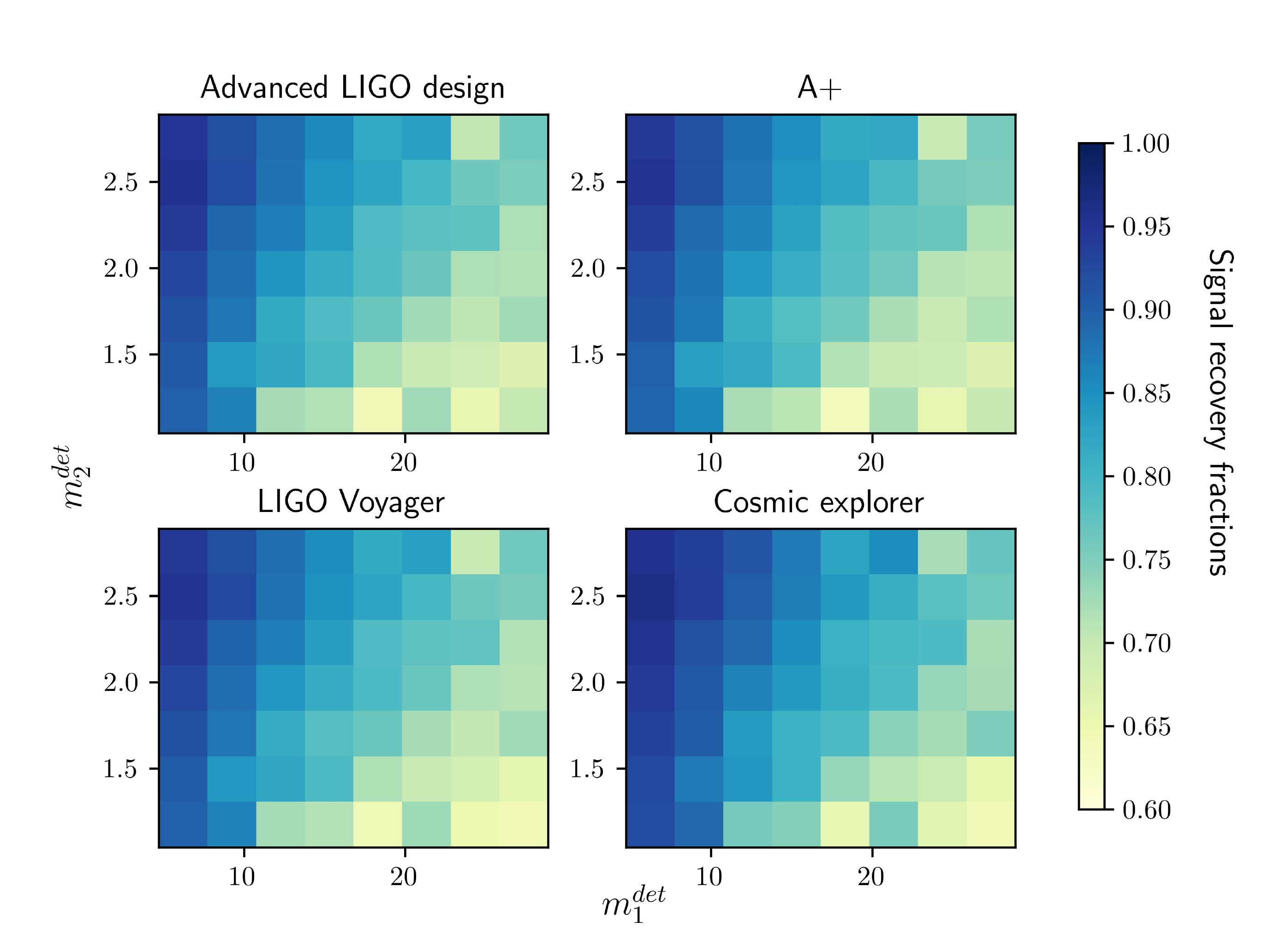}
        \caption{The fraction of detectable precessing sources (with HMs) at a fixed SNR threshold (SRF) $m_1^{det}-m_2^{det}$ for different detector configurations. We observe a similar trend as the FF distribution in Fig.~\ref{fig:FF_diffPSDs.png}; the SRF decreases as the mass-ratio increases and there is no significant change in sensitivity across various detector configurations. We observe a loss of up to $40 \%$ for precessing sources ($13\%$ for non-precessing sources) with higher modes included.}
        \label{fig:SRF_diffPSDs}
\end{figure*}

In the previous section, we have studied the distribution of SNR fraction recovered by an aligned-spin search with respect to an idealized search that could recover all of a signal's SNR. 
A real search will have an observational bias towards signals which are louder in comparison to others, which is not taken into account in the FF distribution. We can estimate the fraction of signals (SRF) that would instead be detected by an aligned-spin search w.r.t. an idealized search assuming a fixed SNR detection threshold.
In Fig. \ref{fig:SRF_diffPSDs} we show the distribution of the SRF for precessing signals with HMs as a function of $(m_1^{det}, m_2^{det})$ for all four detector configurations. We notice a similar trend as the FF distribution; decreasing SRF with increasing mass-ratio of the binary. We find dominant mode aligned-spin template banks will miss up to $\sim 13\%$ of nonprecessing signals with HMs (not shown in the figure) and up to $\sim 40\%$ of the precessing systems assuming our fiducial population. These results hold for each detector configuration.

In practice, a real search that could incorporate the effects of precession and higher-order modes
would incur a trials factor relative to the aligned-spin search; this has subdominant impact on the achievable sensitivity. To give an idea of the magnitude of this effect, we can approximate the relative sensitivity at a fixed false alarm rate, rather than at a fixed SNR threshold (see Eq. (\ref{eq:relative_sensitivity})). As a rough estimate, we take the appropriate $\rho_{\text{thresh}}$, chosen to correspond to a given false alarm rate, from previous works which study developing HM~\cite{Harry:2017weg} and precessing searches~\cite{Harry:2016ijz} as shown in the Table \ref{tab:FAR}. 
Typically less than 1 in every 100 or 1000 years are used to claim a significant event \cite{Usman:2015kfa, Messick:2016aqy}. We evaluate the relative sensitivity $\beta$ at these FARs of an aligned-spin search w.r.t. three different idealized searches: a HM search, a precessing, dominant-mode search, and a precessing higher-mode search. Since, precession effects dominate, we use the same effect threshold for the precessing cases.

\begin{figure*}[h]
        \centering
        \includegraphics[width=13cm, height=10cm]{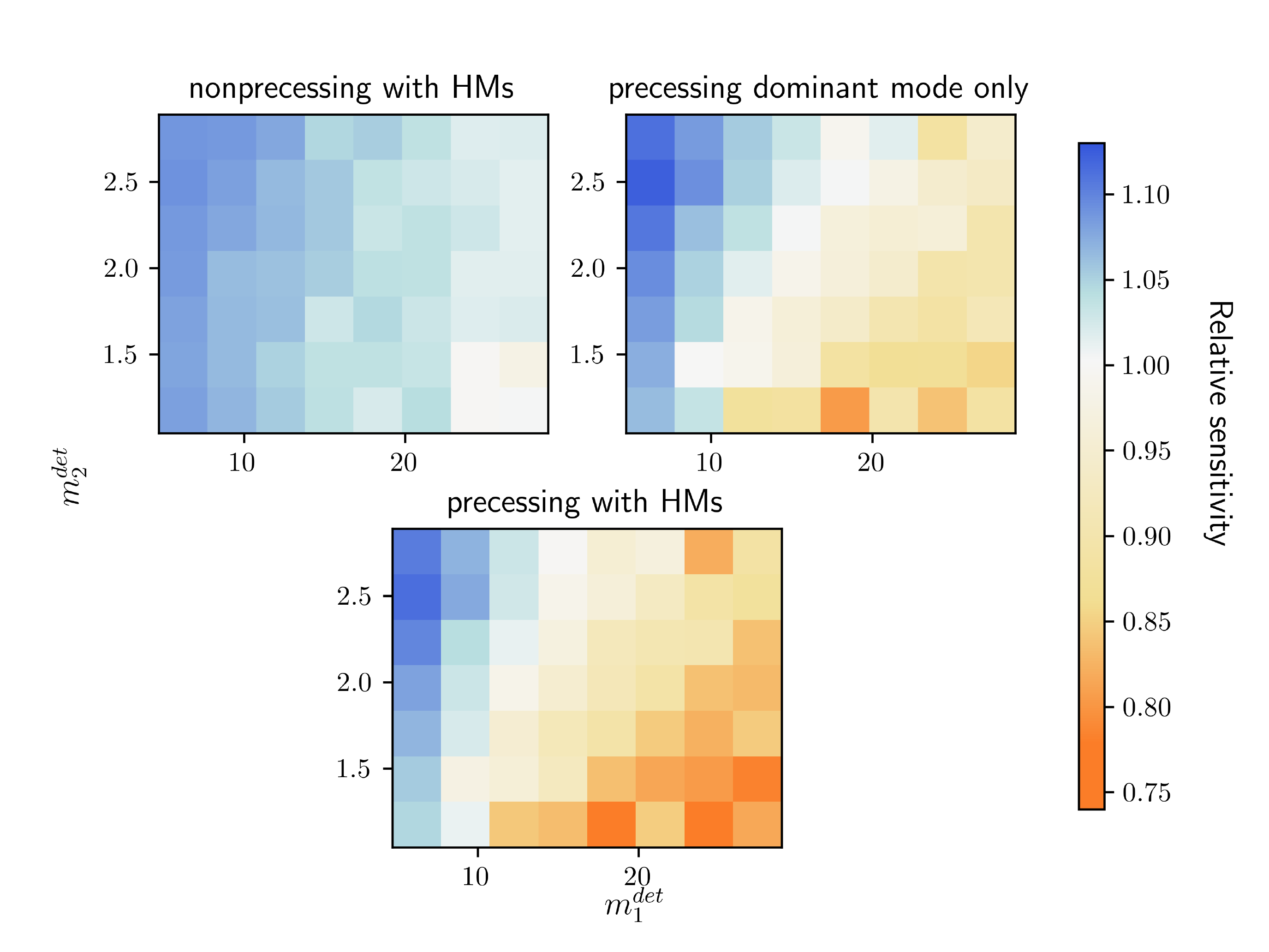}        \caption{The fraction of precessing sources with HMs detected at a fixed false alarm rate relative to an idealized HM (upper left), dominant-mode precessing (upper right), or higher-mode precessing (lower) search as a function of the detector frame component masses $(m_1^{det}, m_2^{det})$. 
        We approximate the increase in background for the reference ideal searches by taking the relevant thresholds from \cite{Harry:2017weg, Harry:2016ijz} at a fixed FAR (as shown in table \ref{tab:FAR}). Regions with values $\geq 1$ indicate better sensitivity of a dominant mode, aligned-spin search relative to a search including additional effects, whereas, regions with modified SRF $< 1$ indicate lower relative sensitivity. We observe that a dominant mode, aligned spin search has better sensitivity than the naive idealized non-precessing search that includes HMs due to the increase in background; however the HM search will have higher sensitivity to edge-on sources. In the regions with $q > 6$, precessing searches will detect up to $\sim 20\%$ more signals for our fiducial signal
        population.}
        \label{fig:aLIGO_weightedSRF}
\end{figure*}

\begin{figure*}[h]
        \centering
        \includegraphics[width=\linewidth]{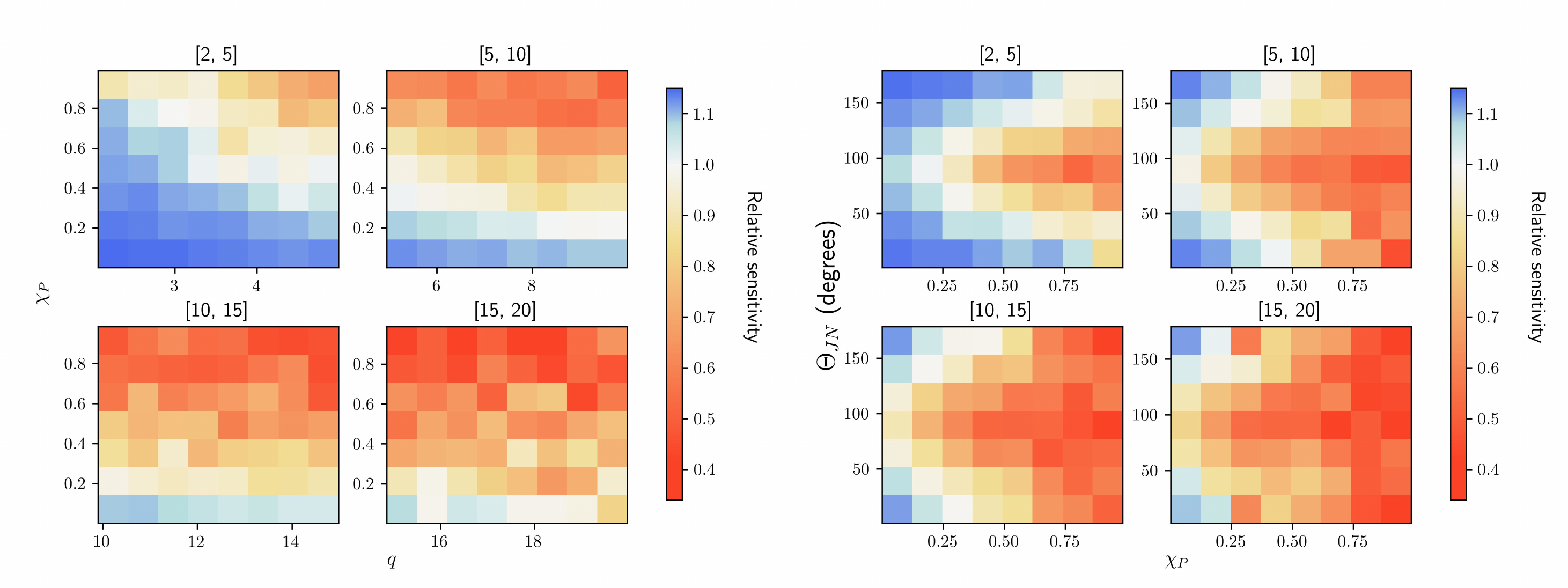}
        \caption{Relative sensitivity of an aligned-spin search w.r.t an idealized precessing search with HMs as a function of  $(q, \chi_P)$ (left) and $(\chi_{P}, \Theta_{JN})$ (right). Small mass-ratio $q \leq 5$ systems which are highly precessing will be missed by a dominant mode, aligned spin searches. For high mass-ratios $q > 6$, there will be a significant loss in sensitivity (up to $60\%$) even for moderately precessing systems.}
        \label{fig:merged}
\end{figure*}

In Fig. \ref{fig:aLIGO_weightedSRF}, we show the sensitivity relative to all three types of searches using the Advanced LIGO design noise curve as a function of ($m_1, m_2$). Parameter regions where the relative sensitivity is $\leq 1$ depicts regions where aligned-spin search is less sensitive than the more advanced, ideal search. When searching for nonprecessing binaries, the aligned-spin bank is more sensitive than a naive search including HMs on average. But, an aligned-spin search loses up to $\sim 25\%$ of highly precessing ($\chi_P > 0.5$) and highly asymmetric ($q > 6$) NSBH binaries averaged over our fiducial population. We further classify sources based on separate mass-ratio regions, with $q\in$ [2, 5], [5, 10], [10, 15] or [15, 20], and plot the relative sensitivity for all the sub-populations as a function of ($q, \chi_p$) (left) and ($\chi_p, \theta_{JN}$) (right) in Fig. \ref{fig:merged}. For low mass-ratio ($q < 5$) binaries we find $\sim 20-30\%$ of highly precessing systems ($\chi_P > 0.5$) will be missed by the aligned-spin search. On the other hand, in the high mass-ratio regions there is loss of $\sim 40-60\%$ ($\sim 30-40\%$) highly (moderately) precessing systems. Aligned-spin searches will also lose $\sim 10-40\%$ low mass-ratio binaries which are nearly edge-on. For high mass-ratios, we find a loss of $\sim 40-65\%$ ($\sim 20-40\%$)  highly (moderately) precessing binaries. These results demonstrate a significant bias against highly precessing, or inclined systems; the development of a practical precessing search would increase the sensitivity to these sources.

\begin{table}[]
    \centering
    \begin{tabular}{c|c|c|c}
         Bank &  FAR $(yr^{-1})$ & $\rho_{\text{ref}}$ & $\rho_{\text{thresh}}$ \\
        \hline
        \hline
        HMs & $0.5 \times 10^{-3}$ & 9.37 &  9.7 \\
        \hline 
        Precession & $0.5 \times 10^{-2}$ & 9.92 & 10.44 \\
        \hline
        HMs with precession & $0.5 \times 10^{-2}$ & 9.92 & 10.44
    \end{tabular}
    \caption{ SNR thresholds ($\rho_{ref}$) (column IV) for different idealized searches at a fixed false alarm rate (II). For the same FAR, the aligned spin searches will require reference SNR thresholds $\rho_{ref}$ (column III). The thresholds are taken from \cite{Harry:2017weg} for an idealized HM search and from \cite{Harry:2016ijz} for a precessing search.}
    \label{tab:FAR}
\end{table}

\section{Developing a Fully Precessing Search}\label{sec:challenges}

Our results indicate a necessity to implement a precessing search for the identified regions of poor sensitivity. One way of implementing a precessing search could be to use the same statistic as Eq. (\ref{Eq:simplified_statistics}) and employ precessing templates. This approximation may not be valid while searching for signals with HMs or precession due to several reasons. For precessing signals, the two polarizations are not related by an overall phase shift $h_+ \not\propto i h_{\times}$, which is what enables the analytic maximization over extrinsic parameters. This is also true in the case of searching using HMs -- as each mode has a unique phase dependence $\Phi_{l,m}$. In the case of a precessing, higher-mode search, both the orbital phase and the inclination of binary, which are typically considered extrinsic parameters, would now behave as intrinsic parameters. The ideal search would marginalize over all extrinsic and intrinsic parameters coherently for all observing detectors. However, a naive implementation would be significantly more expensive than the aligned-spin search due to the increase in the number of intrinsic parameters (spin vectors, orbital phase, inclination), and the need to numerically marginalize over the extrinsic parameters. 

\begin{figure}[H]
    \centering
    \includegraphics[width=\linewidth]{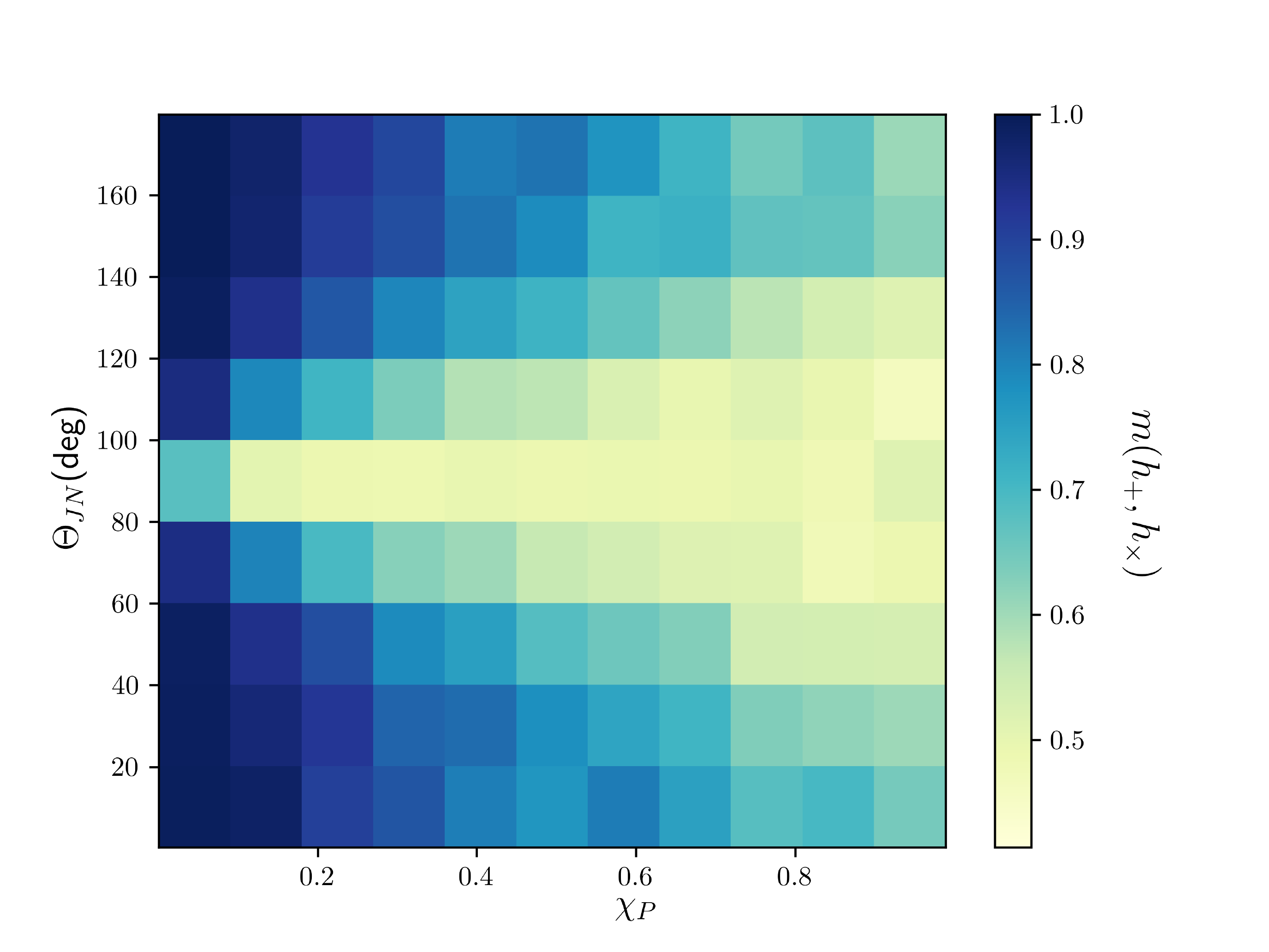}
    \caption{ Distribution of match (overlap maximized over time and phase) between the two polarizations $m(h_{+},h_{\times})$ for a simulated population of precessing signals with higher modes across the two-dimensional surface $(\chi_P, \theta_{JN})$. The values correspond to the fractional loss in SNR measured using a single polarization matched filter statistic; regions with significantly lower values may require a generic statistic.} 
    \label{fig:hplus_hcross}
\end{figure}

We can demonstrate the issue by comparing the match between the two polarizations $m(h_+, h_{\times})$ for every source in our population using precessing signals with HMs. The value (1 - $m(h_+, h_{\times})$) corresponds to the fraction of observed signal SNR lost as a result of approximation in Eq. (\ref{Eq:simplified_statistics}); a value of $m(h_+, h_{\times})) = 0.84$ corresponds to a loss of $16\%$. In Fig. \ref{fig:hplus_hcross}, we show the binned averaged value of the match as a function of ($\chi_P, \theta_{JN}$) at $f_{ref} = 100$ Hz. From the plot, we observe the standard analytical maximization of extrinsic parameters is valid for searching non-precessing systems except when they are nearly edge-on. However, when searching for precessing systems the match decreases with increasing $\chi_P$ or $\theta_{JN}$. We measure a loss in SNR up to $\sim 40-53\%$ for nearly edge-on and up to $\sim 10-40\%$ for face-on/off precessing binaries. These results clearly invalidate the approximation in Eq. (\ref{Eq:simplified_statistics}) and motivate the development of methods to efficiently marginalize over the extrinsic parameters of precessing sources.

Works have proposed new ways of approximating the optimal search statistic~\cite{Harry:2016ijz, Harry:2017weg}. These search methods maximise the SNR over fewer extrinsic parameters than in Eq. (\ref{Eq:simplified_statistics}) and across the remaining ones using a template bank with additional parameters; they also typically do this maximization incoherently between detectors. A generic approach to search for signals with HMs was introduced in \cite{Harry:2017weg}; the search maximised the SNR over ($d_L, \alpha, \delta, \psi$) and uses two additional parameters ($i, \varphi_c$) in the template bank. The same approach is employed in the first search for intermediate-mass BHs including HMs \cite{Chandra:2022ixv}. Similarly,~\cite{Harry:2016ijz} developed an approach to search for precessing signals; their statistic maximises the SNR over sky-parameters and imparts only one additional parameter to the template bank.

\section{Conclusions and discussions}
In this work we have studied the sensitivity of dominant mode, aligned-spin searches, as have been employed in past gravitational-wave surveys ~\cite{LIGOScientific:2021djp, Olsen:2022pin, Nitz:2021uxj}, to a population of NSBH sources. We compare the number of sources an aligned spin search would detect at a fixed false alarm rate relative to idealized searches that incorporate either higher modes, precession, or both. In this study, we've used an estimate of the relative background of each search to account for the change in SNR threshold at a fixed false alarm rate; this could be further improved by a detailed study of precessing template bank generation, which we leave for future work.

We have modelled gravitational-wave signals using a recent waveform model which includes both the effects of precession and higher-order modes~\cite{Pratten:2020ceb}. The model has the largest potential systematics for highly precessing and asymmetrical-mass sources. While we do not expect qualitative changes to our results, improved models would help minimize potential systematic biases in precessing searches.

We find that the aligned spin searches are least effective in detecting sources with asymmetrical masses, large in-plane spins, or binaries with orientation close to edge-on. Dominant mode, aligned-spin searches lose up to $25\%$ of sources with mass-ratios $q > 6$ and up to $60\%$ of highly precessing systems $\chi_p > 0.5$. We compare four different noise curves -- Advanced LIGO design, A+, LIGO Voyager and Cosmic explorer and find that these results apply for each detector configuration we considered. 

Real detector data contains non-Gaussian transient noise~\cite{LIGOScientific:2016gtq, LIGO:2021ppb}; the impact of these artefacts is mitigated using a variety of vetoing methods~\cite{LIGOScientific:2017tza}, including signal consistency tests~\cite{Allen:2004gu, LIGOScientific:2017tza}. In this work we have not considered the impact of these vetoing methods. Since signal-consistency tests compare the expected signal to the data, it is possible that aligned-spin searches could misclassify signals with strong precession or HMs as noise. Therefore, we expect that in practice aligned-spin searches will have even stronger observational biases than presented here.

The detection of precessing systems or signals with higher-order modes provides better constraints of the component spins that carries signatures of the formation of compact-binary sources~\cite{Gompertz:2021xub, Stevenson:2017dlk, Talbot:2017yur, Rodriguez:2016vmx, Johnson-McDaniel:2021rvv}. We have identified regions of parameter space with significant bias against highly precessing systems. The poor sensitivity in these regions motivates the need to develop a search that can fully account for the gravitational-wave signal produced by precessing binaries. Precessing searches can pose computational challenges ~\cite{Harry:2016ijz} which might be overcome by implementing hierarchical strategies~\cite{Dhurkunde:2021csz, Soni:2021vls}. We expect that if we can solve these challenges, our results show that the solution will likely still apply for future detector configurations. Therefore, we strongly recommend a targeted precessing search in the identified regions with poor sensitivity.
 
\acknowledgments
We would like to acknowledge Marlin Sch{\"a}fer for reading the manuscript and for providing useful comments. We acknowledge the Max Planck Gesellschaft and the Atlas cluster computing team at Albert-Einstein Institute (AEI) Hannover for support.
\clearpage

\bibliography{references}

\begin{thebibliography}{75}%
\makeatletter
\providecommand \@ifxundefined [1]{%
 \@ifx{#1\undefined}
}%
\providecommand \@ifnum [1]{%
 \ifnum #1\expandafter \@firstoftwo
 \else \expandafter \@secondoftwo
 \fi
}%
\providecommand \@ifx [1]{%
 \ifx #1\expandafter \@firstoftwo
 \else \expandafter \@secondoftwo
 \fi
}%
\providecommand \natexlab [1]{#1}%
\providecommand \enquote  [1]{``#1''}%
\providecommand \bibnamefont  [1]{#1}%
\providecommand \bibfnamefont [1]{#1}%
\providecommand \citenamefont [1]{#1}%
\providecommand \href@noop [0]{\@secondoftwo}%
\providecommand \href [0]{\begingroup \@sanitize@url \@href}%
\providecommand \@href[1]{\@@startlink{#1}\@@href}%
\providecommand \@@href[1]{\endgroup#1\@@endlink}%
\providecommand \@sanitize@url [0]{\catcode `\\12\catcode `\$12\catcode
  `\&12\catcode `\#12\catcode `\^12\catcode `\_12\catcode `\%12\relax}%
\providecommand \@@startlink[1]{}%
\providecommand \@@endlink[0]{}%
\providecommand \url  [0]{\begingroup\@sanitize@url \@url }%
\providecommand \@url [1]{\endgroup\@href {#1}{\urlprefix }}%
\providecommand \urlprefix  [0]{URL }%
\providecommand \Eprint [0]{\href }%
\providecommand \doibase [0]{http://dx.doi.org/}%
\providecommand \selectlanguage [0]{\@gobble}%
\providecommand \bibinfo  [0]{\@secondoftwo}%
\providecommand \bibfield  [0]{\@secondoftwo}%
\providecommand \translation [1]{[#1]}%
\providecommand \BibitemOpen [0]{}%
\providecommand \bibitemStop [0]{}%
\providecommand \bibitemNoStop [0]{.\EOS\space}%
\providecommand \EOS [0]{\spacefactor3000\relax}%
\providecommand \BibitemShut  [1]{\csname bibitem#1\endcsname}%
\let\auto@bib@innerbib\@empty
\bibitem [{\citenamefont {Abbott}\ \emph
  {et~al.}(2021{\natexlab{a}})\citenamefont {Abbott} \emph
  {et~al.}}]{LIGOScientific:2021djp}%
  \BibitemOpen
  \bibfield  {author} {\bibinfo {author} {\bibfnamefont {R.}~\bibnamefont
  {Abbott}} \emph {et~al.} (\bibinfo {collaboration} {LIGO Scientific, VIRGO,
  KAGRA}),\ }\bibfield  {title} {\enquote {\bibinfo {title} {{GWTC-3: Compact
  Binary Coalescences Observed by LIGO and Virgo During the Second Part of the
  Third Observing Run}},}\ }\href@noop {} {\  (\bibinfo {year}
  {2021}{\natexlab{a}})},\ \Eprint {http://arxiv.org/abs/2111.03606}
  {arXiv:2111.03606 [gr-qc]} \BibitemShut {NoStop}%
\bibitem [{\citenamefont {Nitz}\ \emph {et~al.}(2021)\citenamefont {Nitz},
  \citenamefont {Capano}, \citenamefont {Kumar}, \citenamefont {Wang},
  \citenamefont {Kastha}, \citenamefont {Sch\"afer}, \citenamefont
  {Dhurkunde},\ and\ \citenamefont {Cabero}}]{Nitz:2021uxj}%
  \BibitemOpen
  \bibfield  {author} {\bibinfo {author} {\bibfnamefont {Alexander~H.}\
  \bibnamefont {Nitz}}, \bibinfo {author} {\bibfnamefont {Collin~D.}\
  \bibnamefont {Capano}}, \bibinfo {author} {\bibfnamefont {Sumit}\
  \bibnamefont {Kumar}}, \bibinfo {author} {\bibfnamefont {Yi-Fan}\
  \bibnamefont {Wang}}, \bibinfo {author} {\bibfnamefont {Shilpa}\ \bibnamefont
  {Kastha}}, \bibinfo {author} {\bibfnamefont {Marlin}\ \bibnamefont
  {Sch\"afer}}, \bibinfo {author} {\bibfnamefont {Rahul}\ \bibnamefont
  {Dhurkunde}}, \ and\ \bibinfo {author} {\bibfnamefont {Miriam}\ \bibnamefont
  {Cabero}},\ }\bibfield  {title} {\enquote {\bibinfo {title} {{3-OGC: Catalog
  of Gravitational Waves from Compact-binary Mergers}},}\ }\href {\doibase
  10.3847/1538-4357/ac1c03} {\bibfield  {journal} {\bibinfo  {journal}
  {Astrophys. J.}\ }\textbf {\bibinfo {volume} {922}},\ \bibinfo {pages} {76}
  (\bibinfo {year} {2021})},\ \Eprint {http://arxiv.org/abs/2105.09151}
  {arXiv:2105.09151 [astro-ph.HE]} \BibitemShut {NoStop}%
\bibitem [{\citenamefont {Olsen}\ \emph {et~al.}(2022)\citenamefont {Olsen},
  \citenamefont {Venumadhav}, \citenamefont {Mushkin}, \citenamefont {Roulet},
  \citenamefont {Zackay},\ and\ \citenamefont {Zaldarriaga}}]{Olsen:2022pin}%
  \BibitemOpen
  \bibfield  {author} {\bibinfo {author} {\bibfnamefont {Seth}\ \bibnamefont
  {Olsen}}, \bibinfo {author} {\bibfnamefont {Tejaswi}\ \bibnamefont
  {Venumadhav}}, \bibinfo {author} {\bibfnamefont {Jonathan}\ \bibnamefont
  {Mushkin}}, \bibinfo {author} {\bibfnamefont {Javier}\ \bibnamefont
  {Roulet}}, \bibinfo {author} {\bibfnamefont {Barak}\ \bibnamefont {Zackay}},
  \ and\ \bibinfo {author} {\bibfnamefont {Matias}\ \bibnamefont
  {Zaldarriaga}},\ }\bibfield  {title} {\enquote {\bibinfo {title} {{New binary
  black hole mergers in the LIGO--Virgo O3a data}},}\ }\href@noop {} {\
  (\bibinfo {year} {2022})},\ \Eprint {http://arxiv.org/abs/2201.02252}
  {arXiv:2201.02252 [astro-ph.HE]} \BibitemShut {NoStop}%
\bibitem [{\citenamefont {Aasi}\ \emph {et~al.}(2015)\citenamefont {Aasi} \emph
  {et~al.}}]{LIGOScientific:2014pky}%
  \BibitemOpen
  \bibfield  {author} {\bibinfo {author} {\bibfnamefont {J.}~\bibnamefont
  {Aasi}} \emph {et~al.} (\bibinfo {collaboration} {LIGO Scientific}),\
  }\bibfield  {title} {\enquote {\bibinfo {title} {{Advanced LIGO}},}\ }\href
  {\doibase 10.1088/0264-9381/32/7/074001} {\bibfield  {journal} {\bibinfo
  {journal} {Class. Quant. Grav.}\ }\textbf {\bibinfo {volume} {32}},\ \bibinfo
  {pages} {074001} (\bibinfo {year} {2015})},\ \Eprint
  {http://arxiv.org/abs/1411.4547} {arXiv:1411.4547 [gr-qc]} \BibitemShut
  {NoStop}%
\bibitem [{\citenamefont {Acernese}(2015)}]{Acernese:2015gua}%
  \BibitemOpen
  \bibfield  {author} {\bibinfo {author} {\bibfnamefont {F.}~\bibnamefont
  {Acernese}} (\bibinfo {collaboration} {Virgo}),\ }\bibfield  {title}
  {\enquote {\bibinfo {title} {{The Advanced Virgo detector}},}\ }\href
  {\doibase 10.1088/1742-6596/610/1/012014} {\bibfield  {journal} {\bibinfo
  {journal} {J. Phys. Conf. Ser.}\ }\textbf {\bibinfo {volume} {610}},\
  \bibinfo {pages} {012014} (\bibinfo {year} {2015})}\BibitemShut {NoStop}%
\bibitem [{\citenamefont {Abbott}\ \emph
  {et~al.}(2020{\natexlab{a}})\citenamefont {Abbott} \emph
  {et~al.}}]{LIGOScientific:2020aai}%
  \BibitemOpen
  \bibfield  {author} {\bibinfo {author} {\bibfnamefont {B.~P.}\ \bibnamefont
  {Abbott}} \emph {et~al.} (\bibinfo {collaboration} {LIGO Scientific,
  Virgo}),\ }\bibfield  {title} {\enquote {\bibinfo {title} {{GW190425:
  Observation of a Compact Binary Coalescence with Total Mass $\sim 3.4
  M_{\odot}$}},}\ }\href {\doibase 10.3847/2041-8213/ab75f5} {\bibfield
  {journal} {\bibinfo  {journal} {Astrophys. J. Lett.}\ }\textbf {\bibinfo
  {volume} {892}},\ \bibinfo {pages} {L3} (\bibinfo {year}
  {2020}{\natexlab{a}})},\ \Eprint {http://arxiv.org/abs/2001.01761}
  {arXiv:2001.01761 [astro-ph.HE]} \BibitemShut {NoStop}%
\bibitem [{\citenamefont {Abbott}\ \emph
  {et~al.}(2021{\natexlab{b}})\citenamefont {Abbott} \emph
  {et~al.}}]{LIGOScientific:2021qlt}%
  \BibitemOpen
  \bibfield  {author} {\bibinfo {author} {\bibfnamefont {R.}~\bibnamefont
  {Abbott}} \emph {et~al.} (\bibinfo {collaboration} {LIGO Scientific, KAGRA,
  VIRGO}),\ }\bibfield  {title} {\enquote {\bibinfo {title} {{Observation of
  Gravitational Waves from Two Neutron Star\textendash{}Black Hole
  Coalescences}},}\ }\href {\doibase 10.3847/2041-8213/ac082e} {\bibfield
  {journal} {\bibinfo  {journal} {Astrophys. J. Lett.}\ }\textbf {\bibinfo
  {volume} {915}},\ \bibinfo {pages} {L5} (\bibinfo {year}
  {2021}{\natexlab{b}})},\ \Eprint {http://arxiv.org/abs/2106.15163}
  {arXiv:2106.15163 [astro-ph.HE]} \BibitemShut {NoStop}%
\bibitem [{\citenamefont {Margalit}\ and\ \citenamefont
  {Metzger}(2017)}]{Margalit:2017dij}%
  \BibitemOpen
  \bibfield  {author} {\bibinfo {author} {\bibfnamefont {Ben}\ \bibnamefont
  {Margalit}}\ and\ \bibinfo {author} {\bibfnamefont {Brian~D.}\ \bibnamefont
  {Metzger}},\ }\bibfield  {title} {\enquote {\bibinfo {title} {{Constraining
  the Maximum Mass of Neutron Stars From Multi-Messenger Observations of
  GW170817}},}\ }\href {\doibase 10.3847/2041-8213/aa991c} {\bibfield
  {journal} {\bibinfo  {journal} {Astrophys. J. Lett.}\ }\textbf {\bibinfo
  {volume} {850}},\ \bibinfo {pages} {L19} (\bibinfo {year} {2017})},\ \Eprint
  {http://arxiv.org/abs/1710.05938} {arXiv:1710.05938 [astro-ph.HE]}
  \BibitemShut {NoStop}%
\bibitem [{\citenamefont {Abbott}\ \emph
  {et~al.}(2019{\natexlab{a}})\citenamefont {Abbott} \emph
  {et~al.}}]{LIGOScientific:2018jsj}%
  \BibitemOpen
  \bibfield  {author} {\bibinfo {author} {\bibfnamefont {B.~P.}\ \bibnamefont
  {Abbott}} \emph {et~al.} (\bibinfo {collaboration} {LIGO Scientific,
  Virgo}),\ }\bibfield  {title} {\enquote {\bibinfo {title} {{Binary Black Hole
  Population Properties Inferred from the First and Second Observing Runs of
  Advanced LIGO and Advanced Virgo}},}\ }\href {\doibase
  10.3847/2041-8213/ab3800} {\bibfield  {journal} {\bibinfo  {journal}
  {Astrophys. J. Lett.}\ }\textbf {\bibinfo {volume} {882}},\ \bibinfo {pages}
  {L24} (\bibinfo {year} {2019}{\natexlab{a}})},\ \Eprint
  {http://arxiv.org/abs/1811.12940} {arXiv:1811.12940 [astro-ph.HE]}
  \BibitemShut {NoStop}%
\bibitem [{\citenamefont {Gompertz}\ \emph {et~al.}(2022)\citenamefont
  {Gompertz}, \citenamefont {Nicholl}, \citenamefont {Schmidt}, \citenamefont
  {Pratten},\ and\ \citenamefont {Vecchio}}]{Gompertz:2021xub}%
  \BibitemOpen
  \bibfield  {author} {\bibinfo {author} {\bibfnamefont {B.~P.}\ \bibnamefont
  {Gompertz}}, \bibinfo {author} {\bibfnamefont {M.}~\bibnamefont {Nicholl}},
  \bibinfo {author} {\bibfnamefont {P.}~\bibnamefont {Schmidt}}, \bibinfo
  {author} {\bibfnamefont {G.}~\bibnamefont {Pratten}}, \ and\ \bibinfo
  {author} {\bibfnamefont {A.}~\bibnamefont {Vecchio}},\ }\bibfield  {title}
  {\enquote {\bibinfo {title} {{Constraints on compact binary merger evolution
  from spin-orbit misalignment in gravitational-wave observations}},}\ }\href
  {\doibase 10.1093/mnras/stac029} {\bibfield  {journal} {\bibinfo  {journal}
  {Mon. Not. Roy. Astron. Soc.}\ }\textbf {\bibinfo {volume} {511}},\ \bibinfo
  {pages} {1454--1461} (\bibinfo {year} {2022})},\ \Eprint
  {http://arxiv.org/abs/2108.10184} {arXiv:2108.10184 [astro-ph.HE]}
  \BibitemShut {NoStop}%
\bibitem [{\citenamefont {Abbott}\ \emph
  {et~al.}(2018{\natexlab{a}})\citenamefont {Abbott} \emph
  {et~al.}}]{KAGRA:2013rdx}%
  \BibitemOpen
  \bibfield  {author} {\bibinfo {author} {\bibfnamefont {B.~P.}\ \bibnamefont
  {Abbott}} \emph {et~al.} (\bibinfo {collaboration} {KAGRA, LIGO Scientific,
  Virgo, VIRGO}),\ }\bibfield  {title} {\enquote {\bibinfo {title} {{Prospects
  for observing and localizing gravitational-wave transients with Advanced
  LIGO, Advanced Virgo and KAGRA}},}\ }\href {\doibase
  10.1007/s41114-020-00026-9} {\bibfield  {journal} {\bibinfo  {journal}
  {Living Rev. Rel.}\ }\textbf {\bibinfo {volume} {21}},\ \bibinfo {pages} {3}
  (\bibinfo {year} {2018}{\natexlab{a}})},\ \Eprint
  {http://arxiv.org/abs/1304.0670} {arXiv:1304.0670 [gr-qc]} \BibitemShut
  {NoStop}%
\bibitem [{\citenamefont {Cahillane}\ and\ \citenamefont
  {Mansell}(2022)}]{Cahillane:2022pqm}%
  \BibitemOpen
  \bibfield  {author} {\bibinfo {author} {\bibfnamefont {Craig}\ \bibnamefont
  {Cahillane}}\ and\ \bibinfo {author} {\bibfnamefont {Georgia}\ \bibnamefont
  {Mansell}},\ }\bibfield  {title} {\enquote {\bibinfo {title} {{Review of the
  Advanced LIGO Gravitational Wave Observatories Leading to Observing Run
  Four}},}\ }\href {\doibase 10.3390/galaxies10010036} {\bibfield  {journal}
  {\bibinfo  {journal} {Galaxies}\ }\textbf {\bibinfo {volume} {10}},\ \bibinfo
  {pages} {36} (\bibinfo {year} {2022})},\ \Eprint
  {http://arxiv.org/abs/2202.00847} {arXiv:2202.00847 [gr-qc]} \BibitemShut
  {NoStop}%
\bibitem [{\citenamefont {Adhikari}\ \emph {et~al.}(2020)\citenamefont
  {Adhikari} \emph {et~al.}}]{LIGO:2020xsf}%
  \BibitemOpen
  \bibfield  {author} {\bibinfo {author} {\bibfnamefont {R.~X.}\ \bibnamefont
  {Adhikari}} \emph {et~al.} (\bibinfo {collaboration} {LIGO}),\ }\bibfield
  {title} {\enquote {\bibinfo {title} {{A cryogenic silicon interferometer for
  gravitational-wave detection}},}\ }\href {\doibase 10.1088/1361-6382/ab9143}
  {\bibfield  {journal} {\bibinfo  {journal} {Class. Quant. Grav.}\ }\textbf
  {\bibinfo {volume} {37}},\ \bibinfo {pages} {165003} (\bibinfo {year}
  {2020})},\ \Eprint {http://arxiv.org/abs/2001.11173} {arXiv:2001.11173
  [astro-ph.IM]} \BibitemShut {NoStop}%
\bibitem [{\citenamefont {Reitze}\ \emph {et~al.}(2019)\citenamefont {Reitze}
  \emph {et~al.}}]{Reitze:2019iox}%
  \BibitemOpen
  \bibfield  {author} {\bibinfo {author} {\bibfnamefont {David}\ \bibnamefont
  {Reitze}} \emph {et~al.},\ }\bibfield  {title} {\enquote {\bibinfo {title}
  {{Cosmic Explorer: The U.S. Contribution to Gravitational-Wave Astronomy
  beyond LIGO}},}\ }\href@noop {} {\bibfield  {journal} {\bibinfo  {journal}
  {Bull. Am. Astron. Soc.}\ }\textbf {\bibinfo {volume} {51}},\ \bibinfo
  {pages} {035} (\bibinfo {year} {2019})},\ \Eprint
  {http://arxiv.org/abs/1907.04833} {arXiv:1907.04833 [astro-ph.IM]}
  \BibitemShut {NoStop}%
\bibitem [{\citenamefont {Belczynski}\ \emph {et~al.}(2016)\citenamefont
  {Belczynski}, \citenamefont {Repetto}, \citenamefont {Holz}, \citenamefont
  {O'Shaughnessy}, \citenamefont {Bulik}, \citenamefont {Berti}, \citenamefont
  {Fryer},\ and\ \citenamefont {Dominik}}]{Belczynski:2015tba}%
  \BibitemOpen
  \bibfield  {author} {\bibinfo {author} {\bibfnamefont {Krzysztof}\
  \bibnamefont {Belczynski}}, \bibinfo {author} {\bibfnamefont {Serena}\
  \bibnamefont {Repetto}}, \bibinfo {author} {\bibfnamefont {Daniel~E.}\
  \bibnamefont {Holz}}, \bibinfo {author} {\bibfnamefont {Richard}\
  \bibnamefont {O'Shaughnessy}}, \bibinfo {author} {\bibfnamefont {Tomasz}\
  \bibnamefont {Bulik}}, \bibinfo {author} {\bibfnamefont {Emanuele}\
  \bibnamefont {Berti}}, \bibinfo {author} {\bibfnamefont {Christopher}\
  \bibnamefont {Fryer}}, \ and\ \bibinfo {author} {\bibfnamefont {Michal}\
  \bibnamefont {Dominik}},\ }\bibfield  {title} {\enquote {\bibinfo {title}
  {{Compact Binary Merger Rates: Comparison with LIGO/Virgo Upper Limits}},}\
  }\href {\doibase 10.3847/0004-637X/819/2/108} {\bibfield  {journal} {\bibinfo
   {journal} {Astrophys. J.}\ }\textbf {\bibinfo {volume} {819}},\ \bibinfo
  {pages} {108} (\bibinfo {year} {2016})},\ \Eprint
  {http://arxiv.org/abs/1510.04615} {arXiv:1510.04615 [astro-ph.HE]}
  \BibitemShut {NoStop}%
\bibitem [{\citenamefont {Abbott}\ \emph
  {et~al.}(2016{\natexlab{a}})\citenamefont {Abbott} \emph
  {et~al.}}]{LIGOScientific:2016vpg}%
  \BibitemOpen
  \bibfield  {author} {\bibinfo {author} {\bibfnamefont {B.~P.}\ \bibnamefont
  {Abbott}} \emph {et~al.} (\bibinfo {collaboration} {LIGO Scientific,
  Virgo}),\ }\bibfield  {title} {\enquote {\bibinfo {title} {{Astrophysical
  Implications of the Binary Black-Hole Merger GW150914}},}\ }\href {\doibase
  10.3847/2041-8205/818/2/L22} {\bibfield  {journal} {\bibinfo  {journal}
  {Astrophys. J. Lett.}\ }\textbf {\bibinfo {volume} {818}},\ \bibinfo {pages}
  {L22} (\bibinfo {year} {2016}{\natexlab{a}})},\ \Eprint
  {http://arxiv.org/abs/1602.03846} {arXiv:1602.03846 [astro-ph.HE]}
  \BibitemShut {NoStop}%
\bibitem [{\citenamefont {Broekgaarden}\ and\ \citenamefont
  {Berger}(2021)}]{Broekgaarden:2021hlu}%
  \BibitemOpen
  \bibfield  {author} {\bibinfo {author} {\bibfnamefont {Floor~S.}\
  \bibnamefont {Broekgaarden}}\ and\ \bibinfo {author} {\bibfnamefont {Edo}\
  \bibnamefont {Berger}},\ }\bibfield  {title} {\enquote {\bibinfo {title}
  {{Formation of the First Two Black Hole\textendash{}Neutron Star Mergers
  (GW200115 and GW200105) from Isolated Binary Evolution}},}\ }\href {\doibase
  10.3847/2041-8213/ac2832} {\bibfield  {journal} {\bibinfo  {journal}
  {Astrophys. J. Lett.}\ }\textbf {\bibinfo {volume} {920}},\ \bibinfo {pages}
  {L13} (\bibinfo {year} {2021})},\ \Eprint {http://arxiv.org/abs/2108.05763}
  {arXiv:2108.05763 [astro-ph.HE]} \BibitemShut {NoStop}%
\bibitem [{\citenamefont {Belczynski}\ \emph
  {et~al.}(2010{\natexlab{a}})\citenamefont {Belczynski}, \citenamefont
  {Dominik}, \citenamefont {Bulik}, \citenamefont {O'Shaughnessy},
  \citenamefont {Fryer},\ and\ \citenamefont {Holz}}]{Belczynski:2010tb}%
  \BibitemOpen
  \bibfield  {author} {\bibinfo {author} {\bibfnamefont {Krzysztof}\
  \bibnamefont {Belczynski}}, \bibinfo {author} {\bibfnamefont {Michal}\
  \bibnamefont {Dominik}}, \bibinfo {author} {\bibfnamefont {Tomasz}\
  \bibnamefont {Bulik}}, \bibinfo {author} {\bibfnamefont {Richard}\
  \bibnamefont {O'Shaughnessy}}, \bibinfo {author} {\bibfnamefont {Chris}\
  \bibnamefont {Fryer}}, \ and\ \bibinfo {author} {\bibfnamefont {Daniel~E.}\
  \bibnamefont {Holz}},\ }\bibfield  {title} {\enquote {\bibinfo {title} {{The
  effect of metallicity on the detection prospects for gravitational waves}},}\
  }\href {\doibase 10.1088/2041-8205/715/2/L138} {\bibfield  {journal}
  {\bibinfo  {journal} {Astrophys. J. Lett.}\ }\textbf {\bibinfo {volume}
  {715}},\ \bibinfo {pages} {L138} (\bibinfo {year} {2010}{\natexlab{a}})},\
  \Eprint {http://arxiv.org/abs/1004.0386} {arXiv:1004.0386 [astro-ph.HE]}
  \BibitemShut {NoStop}%
\bibitem [{\citenamefont {Dominik}\ \emph {et~al.}(2012)\citenamefont
  {Dominik}, \citenamefont {Belczynski}, \citenamefont {Fryer}, \citenamefont
  {Holz}, \citenamefont {Berti}, \citenamefont {Bulik}, \citenamefont
  {Mandel},\ and\ \citenamefont {O'Shaughnessy}}]{Dominik:2012kk}%
  \BibitemOpen
  \bibfield  {author} {\bibinfo {author} {\bibfnamefont {Michal}\ \bibnamefont
  {Dominik}}, \bibinfo {author} {\bibfnamefont {Krzysztof}\ \bibnamefont
  {Belczynski}}, \bibinfo {author} {\bibfnamefont {Christopher}\ \bibnamefont
  {Fryer}}, \bibinfo {author} {\bibfnamefont {Daniel}\ \bibnamefont {Holz}},
  \bibinfo {author} {\bibfnamefont {Emanuele}\ \bibnamefont {Berti}}, \bibinfo
  {author} {\bibfnamefont {Tomasz}\ \bibnamefont {Bulik}}, \bibinfo {author}
  {\bibfnamefont {Ilya}\ \bibnamefont {Mandel}}, \ and\ \bibinfo {author}
  {\bibfnamefont {Richard}\ \bibnamefont {O'Shaughnessy}},\ }\bibfield  {title}
  {\enquote {\bibinfo {title} {{Double Compact Objects I: The Significance of
  the Common Envelope on Merger Rates}},}\ }\href {\doibase
  10.1088/0004-637X/759/1/52} {\bibfield  {journal} {\bibinfo  {journal}
  {Astrophys. J.}\ }\textbf {\bibinfo {volume} {759}},\ \bibinfo {pages} {52}
  (\bibinfo {year} {2012})},\ \Eprint {http://arxiv.org/abs/1202.4901}
  {arXiv:1202.4901 [astro-ph.HE]} \BibitemShut {NoStop}%
\bibitem [{\citenamefont {Pooley}\ \emph {et~al.}(2003)\citenamefont {Pooley}
  \emph {et~al.}}]{Pooley:2003zb}%
  \BibitemOpen
  \bibfield  {author} {\bibinfo {author} {\bibfnamefont {D.}~\bibnamefont
  {Pooley}} \emph {et~al.},\ }\bibfield  {title} {\enquote {\bibinfo {title}
  {{Dynamical formation of close binary systems in globular clusters}},}\
  }\href {\doibase 10.1086/377074} {\bibfield  {journal} {\bibinfo  {journal}
  {Astrophys. J. Lett.}\ }\textbf {\bibinfo {volume} {591}},\ \bibinfo {pages}
  {L131--L134} (\bibinfo {year} {2003})},\ \Eprint
  {http://arxiv.org/abs/astro-ph/0305003} {arXiv:astro-ph/0305003} \BibitemShut
  {NoStop}%
\bibitem [{\citenamefont {Rodriguez}\ \emph {et~al.}(2015)\citenamefont
  {Rodriguez}, \citenamefont {Morscher}, \citenamefont {Pattabiraman},
  \citenamefont {Chatterjee}, \citenamefont {Haster},\ and\ \citenamefont
  {Rasio}}]{Rodriguez:2015oxa}%
  \BibitemOpen
  \bibfield  {author} {\bibinfo {author} {\bibfnamefont {Carl~L.}\ \bibnamefont
  {Rodriguez}}, \bibinfo {author} {\bibfnamefont {Meagan}\ \bibnamefont
  {Morscher}}, \bibinfo {author} {\bibfnamefont {Bharath}\ \bibnamefont
  {Pattabiraman}}, \bibinfo {author} {\bibfnamefont {Sourav}\ \bibnamefont
  {Chatterjee}}, \bibinfo {author} {\bibfnamefont {Carl-Johan}\ \bibnamefont
  {Haster}}, \ and\ \bibinfo {author} {\bibfnamefont {Frederic~A.}\
  \bibnamefont {Rasio}},\ }\bibfield  {title} {\enquote {\bibinfo {title}
  {{Binary Black Hole Mergers from Globular Clusters: Implications for Advanced
  LIGO}},}\ }\href {\doibase 10.1103/PhysRevLett.115.051101} {\bibfield
  {journal} {\bibinfo  {journal} {Phys. Rev. Lett.}\ }\textbf {\bibinfo
  {volume} {115}},\ \bibinfo {pages} {051101} (\bibinfo {year} {2015})},\
  \bibinfo {note} {[Erratum: Phys.Rev.Lett. 116, 029901 (2016)]},\ \Eprint
  {http://arxiv.org/abs/1505.00792} {arXiv:1505.00792 [astro-ph.HE]}
  \BibitemShut {NoStop}%
\bibitem [{\citenamefont {Stevenson}\ \emph {et~al.}(2017)\citenamefont
  {Stevenson}, \citenamefont {Berry},\ and\ \citenamefont
  {Mandel}}]{Stevenson:2017dlk}%
  \BibitemOpen
  \bibfield  {author} {\bibinfo {author} {\bibfnamefont {Simon}\ \bibnamefont
  {Stevenson}}, \bibinfo {author} {\bibfnamefont {Christopher P.~L.}\
  \bibnamefont {Berry}}, \ and\ \bibinfo {author} {\bibfnamefont {Ilya}\
  \bibnamefont {Mandel}},\ }\bibfield  {title} {\enquote {\bibinfo {title}
  {{Hierarchical analysis of gravitational-wave measurements of binary black
  hole spin\textendash{}orbit misalignments}},}\ }\href {\doibase
  10.1093/mnras/stx1764} {\bibfield  {journal} {\bibinfo  {journal} {Mon. Not.
  Roy. Astron. Soc.}\ }\textbf {\bibinfo {volume} {471}},\ \bibinfo {pages}
  {2801--2811} (\bibinfo {year} {2017})},\ \Eprint
  {http://arxiv.org/abs/1703.06873} {arXiv:1703.06873 [astro-ph.HE]}
  \BibitemShut {NoStop}%
\bibitem [{\citenamefont {Talbot}\ and\ \citenamefont
  {Thrane}(2017)}]{Talbot:2017yur}%
  \BibitemOpen
  \bibfield  {author} {\bibinfo {author} {\bibfnamefont {Colm}\ \bibnamefont
  {Talbot}}\ and\ \bibinfo {author} {\bibfnamefont {Eric}\ \bibnamefont
  {Thrane}},\ }\bibfield  {title} {\enquote {\bibinfo {title} {{Determining the
  population properties of spinning black holes}},}\ }\href {\doibase
  10.1103/PhysRevD.96.023012} {\bibfield  {journal} {\bibinfo  {journal} {Phys.
  Rev. D}\ }\textbf {\bibinfo {volume} {96}},\ \bibinfo {pages} {023012}
  (\bibinfo {year} {2017})},\ \Eprint {http://arxiv.org/abs/1704.08370}
  {arXiv:1704.08370 [astro-ph.HE]} \BibitemShut {NoStop}%
\bibitem [{\citenamefont {Rodriguez}\ \emph {et~al.}(2016)\citenamefont
  {Rodriguez}, \citenamefont {Zevin}, \citenamefont {Pankow}, \citenamefont
  {Kalogera},\ and\ \citenamefont {Rasio}}]{Rodriguez:2016vmx}%
  \BibitemOpen
  \bibfield  {author} {\bibinfo {author} {\bibfnamefont {Carl~L.}\ \bibnamefont
  {Rodriguez}}, \bibinfo {author} {\bibfnamefont {Michael}\ \bibnamefont
  {Zevin}}, \bibinfo {author} {\bibfnamefont {Chris}\ \bibnamefont {Pankow}},
  \bibinfo {author} {\bibfnamefont {Vasilliki}\ \bibnamefont {Kalogera}}, \
  and\ \bibinfo {author} {\bibfnamefont {Frederic~A.}\ \bibnamefont {Rasio}},\
  }\bibfield  {title} {\enquote {\bibinfo {title} {{Illuminating Black Hole
  Binary Formation Channels with Spins in Advanced LIGO}},}\ }\href {\doibase
  10.3847/2041-8205/832/1/L2} {\bibfield  {journal} {\bibinfo  {journal}
  {Astrophys. J. Lett.}\ }\textbf {\bibinfo {volume} {832}},\ \bibinfo {pages}
  {L2} (\bibinfo {year} {2016})},\ \Eprint {http://arxiv.org/abs/1609.05916}
  {arXiv:1609.05916 [astro-ph.HE]} \BibitemShut {NoStop}%
\bibitem [{\citenamefont {Johnson-McDaniel}\ \emph {et~al.}(2022)\citenamefont
  {Johnson-McDaniel}, \citenamefont {Kulkarni},\ and\ \citenamefont
  {Gupta}}]{Johnson-McDaniel:2021rvv}%
  \BibitemOpen
  \bibfield  {author} {\bibinfo {author} {\bibfnamefont {Nathan~K.}\
  \bibnamefont {Johnson-McDaniel}}, \bibinfo {author} {\bibfnamefont {Sumeet}\
  \bibnamefont {Kulkarni}}, \ and\ \bibinfo {author} {\bibfnamefont {Anuradha}\
  \bibnamefont {Gupta}},\ }\bibfield  {title} {\enquote {\bibinfo {title}
  {{Inferring spin tilts at formation from gravitational wave observations of
  binary black holes: Interfacing precession-averaged and orbit-averaged spin
  evolution}},}\ }\href {\doibase 10.1103/PhysRevD.106.023001} {\bibfield
  {journal} {\bibinfo  {journal} {Phys. Rev. D}\ }\textbf {\bibinfo {volume}
  {106}},\ \bibinfo {pages} {023001} (\bibinfo {year} {2022})},\ \Eprint
  {http://arxiv.org/abs/2107.11902} {arXiv:2107.11902 [astro-ph.HE]}
  \BibitemShut {NoStop}%
\bibitem [{\citenamefont {Krishnendu}\ and\ \citenamefont
  {Ohme}(2022)}]{Krishnendu:2021cyi}%
  \BibitemOpen
  \bibfield  {author} {\bibinfo {author} {\bibfnamefont {N.~V.}\ \bibnamefont
  {Krishnendu}}\ and\ \bibinfo {author} {\bibfnamefont {Frank}\ \bibnamefont
  {Ohme}},\ }\bibfield  {title} {\enquote {\bibinfo {title} {{Interplay of
  spin-precession and higher harmonics in the parameter estimation of binary
  black holes}},}\ }\href {\doibase 10.1103/PhysRevD.105.064012} {\bibfield
  {journal} {\bibinfo  {journal} {Phys. Rev. D}\ }\textbf {\bibinfo {volume}
  {105}},\ \bibinfo {pages} {064012} (\bibinfo {year} {2022})},\ \Eprint
  {http://arxiv.org/abs/2110.00766} {arXiv:2110.00766 [gr-qc]} \BibitemShut
  {NoStop}%
\bibitem [{\citenamefont {Hannam}\ \emph {et~al.}(2021)\citenamefont {Hannam},
  \citenamefont {Hoy}, \citenamefont {Thompson}, \citenamefont {Fairhurst},\
  and\ \citenamefont {Raymond}}]{Hannam:2021pit}%
  \BibitemOpen
  \bibfield  {author} {\bibinfo {author} {\bibfnamefont {Mark}\ \bibnamefont
  {Hannam}}, \bibinfo {author} {\bibfnamefont {Charlie}\ \bibnamefont {Hoy}},
  \bibinfo {author} {\bibfnamefont {Jonathan~E.}\ \bibnamefont {Thompson}},
  \bibinfo {author} {\bibfnamefont {Stephen}\ \bibnamefont {Fairhurst}}, \ and\
  \bibinfo {author} {\bibfnamefont {Vivien}\ \bibnamefont {Raymond}} (\bibinfo
  {collaboration} {VIRGO}),\ }\bibfield  {title} {\enquote {\bibinfo {title}
  {{Measurement of general-relativistic precession in a black-hole binary}},}\
  }\href@noop {} {\  (\bibinfo {year} {2021})},\ \Eprint
  {http://arxiv.org/abs/2112.11300} {arXiv:2112.11300 [gr-qc]} \BibitemShut
  {NoStop}%
\bibitem [{\citenamefont {Abbott}\ \emph
  {et~al.}(2020{\natexlab{b}})\citenamefont {Abbott} \emph
  {et~al.}}]{LIGOScientific:2020stg}%
  \BibitemOpen
  \bibfield  {author} {\bibinfo {author} {\bibfnamefont {R.}~\bibnamefont
  {Abbott}} \emph {et~al.} (\bibinfo {collaboration} {LIGO Scientific,
  Virgo}),\ }\bibfield  {title} {\enquote {\bibinfo {title} {{GW190412:
  Observation of a Binary-Black-Hole Coalescence with Asymmetric Masses}},}\
  }\href {\doibase 10.1103/PhysRevD.102.043015} {\bibfield  {journal} {\bibinfo
   {journal} {Phys. Rev. D}\ }\textbf {\bibinfo {volume} {102}},\ \bibinfo
  {pages} {043015} (\bibinfo {year} {2020}{\natexlab{b}})},\ \Eprint
  {http://arxiv.org/abs/2004.08342} {arXiv:2004.08342 [astro-ph.HE]}
  \BibitemShut {NoStop}%
\bibitem [{\citenamefont {Abbott}\ \emph
  {et~al.}(2020{\natexlab{c}})\citenamefont {Abbott} \emph
  {et~al.}}]{LIGOScientific:2020zkf}%
  \BibitemOpen
  \bibfield  {author} {\bibinfo {author} {\bibfnamefont {R.}~\bibnamefont
  {Abbott}} \emph {et~al.} (\bibinfo {collaboration} {LIGO Scientific,
  Virgo}),\ }\bibfield  {title} {\enquote {\bibinfo {title} {{GW190814:
  Gravitational Waves from the Coalescence of a 23 Solar Mass Black Hole with a
  2.6 Solar Mass Compact Object}},}\ }\href {\doibase 10.3847/2041-8213/ab960f}
  {\bibfield  {journal} {\bibinfo  {journal} {Astrophys. J. Lett.}\ }\textbf
  {\bibinfo {volume} {896}},\ \bibinfo {pages} {L44} (\bibinfo {year}
  {2020}{\natexlab{c}})},\ \Eprint {http://arxiv.org/abs/2006.12611}
  {arXiv:2006.12611 [astro-ph.HE]} \BibitemShut {NoStop}%
\bibitem [{\citenamefont {Capano}\ \emph {et~al.}(2021)\citenamefont {Capano},
  \citenamefont {Cabero}, \citenamefont {Westerweck}, \citenamefont {Abedi},
  \citenamefont {Kastha}, \citenamefont {Nitz}, \citenamefont {Nielsen},\ and\
  \citenamefont {Krishnan}}]{Capano:2021etf}%
  \BibitemOpen
  \bibfield  {author} {\bibinfo {author} {\bibfnamefont {Collin~D.}\
  \bibnamefont {Capano}}, \bibinfo {author} {\bibfnamefont {Miriam}\
  \bibnamefont {Cabero}}, \bibinfo {author} {\bibfnamefont {Julian}\
  \bibnamefont {Westerweck}}, \bibinfo {author} {\bibfnamefont {Jahed}\
  \bibnamefont {Abedi}}, \bibinfo {author} {\bibfnamefont {Shilpa}\
  \bibnamefont {Kastha}}, \bibinfo {author} {\bibfnamefont {Alexander~H.}\
  \bibnamefont {Nitz}}, \bibinfo {author} {\bibfnamefont {Alex~B.}\
  \bibnamefont {Nielsen}}, \ and\ \bibinfo {author} {\bibfnamefont {Badri}\
  \bibnamefont {Krishnan}},\ }\bibfield  {title} {\enquote {\bibinfo {title}
  {{Observation of a multimode quasi-normal spectrum from a perturbed black
  hole}},}\ }\href@noop {} {\  (\bibinfo {year} {2021})},\ \Eprint
  {http://arxiv.org/abs/2105.05238} {arXiv:2105.05238 [gr-qc]} \BibitemShut
  {NoStop}%
\bibitem [{\citenamefont {Broekgaarden}\ \emph
  {et~al.}(2021{\natexlab{a}})\citenamefont {Broekgaarden} \emph
  {et~al.}}]{Broekgaarden:2021efa}%
  \BibitemOpen
  \bibfield  {author} {\bibinfo {author} {\bibfnamefont {Floor~S.}\
  \bibnamefont {Broekgaarden}} \emph {et~al.},\ }\bibfield  {title} {\enquote
  {\bibinfo {title} {{Impact of Massive Binary Star and Cosmic Evolution on
  Gravitational Wave Observations II: Double Compact Object Rates and
  Properties}},}\ }\href@noop {} {\  (\bibinfo {year} {2021}{\natexlab{a}})},\
  \Eprint {http://arxiv.org/abs/2112.05763} {arXiv:2112.05763 [astro-ph.HE]}
  \BibitemShut {NoStop}%
\bibitem [{\citenamefont {Fryer}\ and\ \citenamefont
  {Kalogera}(2001)}]{Fryer:1999ht}%
  \BibitemOpen
  \bibfield  {author} {\bibinfo {author} {\bibfnamefont {Chris~L.}\
  \bibnamefont {Fryer}}\ and\ \bibinfo {author} {\bibfnamefont {Vassiliki}\
  \bibnamefont {Kalogera}},\ }\bibfield  {title} {\enquote {\bibinfo {title}
  {{Theoretical black hole mass distributions}},}\ }\href {\doibase
  10.1086/321359} {\bibfield  {journal} {\bibinfo  {journal} {Astrophys. J.}\
  }\textbf {\bibinfo {volume} {554}},\ \bibinfo {pages} {548--560} (\bibinfo
  {year} {2001})},\ \Eprint {http://arxiv.org/abs/astro-ph/9911312}
  {arXiv:astro-ph/9911312} \BibitemShut {NoStop}%
\bibitem [{\citenamefont {Belczynski}\ \emph
  {et~al.}(2010{\natexlab{b}})\citenamefont {Belczynski}, \citenamefont
  {Bulik}, \citenamefont {Fryer}, \citenamefont {Ruiter}, \citenamefont
  {Vink},\ and\ \citenamefont {Hurley}}]{Belczynski:2009xy}%
  \BibitemOpen
  \bibfield  {author} {\bibinfo {author} {\bibfnamefont {Krzysztof}\
  \bibnamefont {Belczynski}}, \bibinfo {author} {\bibfnamefont {Tomasz}\
  \bibnamefont {Bulik}}, \bibinfo {author} {\bibfnamefont {Chris~L.}\
  \bibnamefont {Fryer}}, \bibinfo {author} {\bibfnamefont {Ashley}\
  \bibnamefont {Ruiter}}, \bibinfo {author} {\bibfnamefont {Jorick~S.}\
  \bibnamefont {Vink}}, \ and\ \bibinfo {author} {\bibfnamefont {Jarrod~R.}\
  \bibnamefont {Hurley}},\ }\bibfield  {title} {\enquote {\bibinfo {title} {{On
  The Maximum Mass of Stellar Black Holes}},}\ }\href {\doibase
  10.1088/0004-637X/714/2/1217} {\bibfield  {journal} {\bibinfo  {journal}
  {Astrophys. J.}\ }\textbf {\bibinfo {volume} {714}},\ \bibinfo {pages}
  {1217--1226} (\bibinfo {year} {2010}{\natexlab{b}})},\ \Eprint
  {http://arxiv.org/abs/0904.2784} {arXiv:0904.2784 [astro-ph.SR]} \BibitemShut
  {NoStop}%
\bibitem [{\citenamefont {Lyne}\ and\ \citenamefont
  {Lorimer}(1994)}]{Lyne:1994az}%
  \BibitemOpen
  \bibfield  {author} {\bibinfo {author} {\bibfnamefont {A.~G.}\ \bibnamefont
  {Lyne}}\ and\ \bibinfo {author} {\bibfnamefont {D.~R.}\ \bibnamefont
  {Lorimer}},\ }\bibfield  {title} {\enquote {\bibinfo {title} {{High birth
  velocities of radio pulsars}},}\ }\href {\doibase 10.1038/369127a0}
  {\bibfield  {journal} {\bibinfo  {journal} {Nature}\ }\textbf {\bibinfo
  {volume} {369}},\ \bibinfo {pages} {127} (\bibinfo {year}
  {1994})}\BibitemShut {NoStop}%
\bibitem [{\citenamefont {Janka}(2013)}]{Janka:2013hfa}%
  \BibitemOpen
  \bibfield  {author} {\bibinfo {author} {\bibfnamefont {H.~Thomas}\
  \bibnamefont {Janka}},\ }\bibfield  {title} {\enquote {\bibinfo {title}
  {{Natal Kicks of Stellar-Mass Black Holes by Asymmetric Mass Ejection in
  Fallback Supernovae}},}\ }\href {\doibase 10.1093/mnras/stt1106} {\bibfield
  {journal} {\bibinfo  {journal} {Mon. Not. Roy. Astron. Soc.}\ }\textbf
  {\bibinfo {volume} {434}},\ \bibinfo {pages} {1355} (\bibinfo {year}
  {2013})},\ \Eprint {http://arxiv.org/abs/1306.0007} {arXiv:1306.0007
  [astro-ph.SR]} \BibitemShut {NoStop}%
\bibitem [{\citenamefont {Broekgaarden}\ \emph
  {et~al.}(2021{\natexlab{b}})\citenamefont {Broekgaarden}, \citenamefont
  {Berger}, \citenamefont {Neijssel}, \citenamefont {Vigna-G\'omez},
  \citenamefont {Chattopadhyay}, \citenamefont {Stevenson}, \citenamefont
  {Chruslinska}, \citenamefont {Justham}, \citenamefont {de~Mink},\ and\
  \citenamefont {Mandel}}]{Broekgaarden:2021iew}%
  \BibitemOpen
  \bibfield  {author} {\bibinfo {author} {\bibfnamefont {Floor~S.}\
  \bibnamefont {Broekgaarden}}, \bibinfo {author} {\bibfnamefont {Edo}\
  \bibnamefont {Berger}}, \bibinfo {author} {\bibfnamefont {Coenraad~J.}\
  \bibnamefont {Neijssel}}, \bibinfo {author} {\bibfnamefont {Alejandro}\
  \bibnamefont {Vigna-G\'omez}}, \bibinfo {author} {\bibfnamefont {Debatri}\
  \bibnamefont {Chattopadhyay}}, \bibinfo {author} {\bibfnamefont {Simon}\
  \bibnamefont {Stevenson}}, \bibinfo {author} {\bibfnamefont {Martyna}\
  \bibnamefont {Chruslinska}}, \bibinfo {author} {\bibfnamefont {Stephen}\
  \bibnamefont {Justham}}, \bibinfo {author} {\bibfnamefont {Selma~E.}\
  \bibnamefont {de~Mink}}, \ and\ \bibinfo {author} {\bibfnamefont {Ilya}\
  \bibnamefont {Mandel}},\ }\bibfield  {title} {\enquote {\bibinfo {title}
  {{Impact of massive binary star and cosmic evolution on gravitational wave
  observations I: black hole\textendash{}neutron star mergers}},}\ }\href
  {\doibase 10.1093/mnras/stab2716} {\bibfield  {journal} {\bibinfo  {journal}
  {Mon. Not. Roy. Astron. Soc.}\ }\textbf {\bibinfo {volume} {508}},\ \bibinfo
  {pages} {5028--5063} (\bibinfo {year} {2021}{\natexlab{b}})},\ \Eprint
  {http://arxiv.org/abs/2103.02608} {arXiv:2103.02608 [astro-ph.HE]}
  \BibitemShut {NoStop}%
\bibitem [{\citenamefont {Gupta}\ \emph {et~al.}(2017)\citenamefont {Gupta},
  \citenamefont {Arun},\ and\ \citenamefont {Sathyaprakash}}]{Gupta:2017dsq}%
  \BibitemOpen
  \bibfield  {author} {\bibinfo {author} {\bibfnamefont {Anuradha}\
  \bibnamefont {Gupta}}, \bibinfo {author} {\bibfnamefont {K.~G.}\ \bibnamefont
  {Arun}}, \ and\ \bibinfo {author} {\bibfnamefont {B.~S.}\ \bibnamefont
  {Sathyaprakash}},\ }\bibfield  {title} {\enquote {\bibinfo {title}
  {{Implications of Binary Black Hole Detections on the Merger Rates of Double
  Neutron Stars and Neutron Star\textendash{}Black Holes}},}\ }\href {\doibase
  10.3847/2041-8213/aa9271} {\bibfield  {journal} {\bibinfo  {journal}
  {Astrophys. J. Lett.}\ }\textbf {\bibinfo {volume} {849}},\ \bibinfo {pages}
  {L14} (\bibinfo {year} {2017})},\ \Eprint {http://arxiv.org/abs/1708.03939}
  {arXiv:1708.03939 [astro-ph.IM]} \BibitemShut {NoStop}%
\bibitem [{\citenamefont {Harry}\ \emph {et~al.}(2014)\citenamefont {Harry},
  \citenamefont {Nitz}, \citenamefont {Brown}, \citenamefont {Lundgren},
  \citenamefont {Ochsner},\ and\ \citenamefont {Keppel}}]{Harry:2013tca}%
  \BibitemOpen
  \bibfield  {author} {\bibinfo {author} {\bibfnamefont {Ian~W.}\ \bibnamefont
  {Harry}}, \bibinfo {author} {\bibfnamefont {Alexander~H.}\ \bibnamefont
  {Nitz}}, \bibinfo {author} {\bibfnamefont {Duncan~A.}\ \bibnamefont {Brown}},
  \bibinfo {author} {\bibfnamefont {Andrew~P.}\ \bibnamefont {Lundgren}},
  \bibinfo {author} {\bibfnamefont {Evan}\ \bibnamefont {Ochsner}}, \ and\
  \bibinfo {author} {\bibfnamefont {Drew}\ \bibnamefont {Keppel}},\ }\bibfield
  {title} {\enquote {\bibinfo {title} {{Investigating the effect of precession
  on searches for neutron-star-black-hole binaries with Advanced LIGO}},}\
  }\href {\doibase 10.1103/PhysRevD.89.024010} {\bibfield  {journal} {\bibinfo
  {journal} {Phys. Rev. D}\ }\textbf {\bibinfo {volume} {89}},\ \bibinfo
  {pages} {024010} (\bibinfo {year} {2014})},\ \Eprint
  {http://arxiv.org/abs/1307.3562} {arXiv:1307.3562 [gr-qc]} \BibitemShut
  {NoStop}%
\bibitem [{\citenamefont {Usman}\ \emph {et~al.}(2016)\citenamefont {Usman}
  \emph {et~al.}}]{Usman:2015kfa}%
  \BibitemOpen
  \bibfield  {author} {\bibinfo {author} {\bibfnamefont {Samantha~A.}\
  \bibnamefont {Usman}} \emph {et~al.},\ }\bibfield  {title} {\enquote
  {\bibinfo {title} {{The PyCBC search for gravitational waves from compact
  binary coalescence}},}\ }\href {\doibase 10.1088/0264-9381/33/21/215004}
  {\bibfield  {journal} {\bibinfo  {journal} {Class. Quant. Grav.}\ }\textbf
  {\bibinfo {volume} {33}},\ \bibinfo {pages} {215004} (\bibinfo {year}
  {2016})},\ \Eprint {http://arxiv.org/abs/1508.02357} {arXiv:1508.02357
  [gr-qc]} \BibitemShut {NoStop}%
\bibitem [{\citenamefont {Messick}\ \emph {et~al.}(2017)\citenamefont {Messick}
  \emph {et~al.}}]{Messick:2016aqy}%
  \BibitemOpen
  \bibfield  {author} {\bibinfo {author} {\bibfnamefont {Cody}\ \bibnamefont
  {Messick}} \emph {et~al.},\ }\bibfield  {title} {\enquote {\bibinfo {title}
  {{Analysis Framework for the Prompt Discovery of Compact Binary Mergers in
  Gravitational-wave Data}},}\ }\href {\doibase 10.1103/PhysRevD.95.042001}
  {\bibfield  {journal} {\bibinfo  {journal} {Phys. Rev. D}\ }\textbf {\bibinfo
  {volume} {95}},\ \bibinfo {pages} {042001} (\bibinfo {year} {2017})},\
  \Eprint {http://arxiv.org/abs/1604.04324} {arXiv:1604.04324 [astro-ph.IM]}
  \BibitemShut {NoStop}%
\bibitem [{\citenamefont {Aubin}\ \emph {et~al.}(2021)\citenamefont {Aubin}
  \emph {et~al.}}]{Aubin:2020goo}%
  \BibitemOpen
  \bibfield  {author} {\bibinfo {author} {\bibfnamefont {F.}~\bibnamefont
  {Aubin}} \emph {et~al.},\ }\bibfield  {title} {\enquote {\bibinfo {title}
  {{The MBTA pipeline for detecting compact binary coalescences in the third
  LIGO\textendash{}Virgo observing run}},}\ }\href {\doibase
  10.1088/1361-6382/abe913} {\bibfield  {journal} {\bibinfo  {journal} {Class.
  Quant. Grav.}\ }\textbf {\bibinfo {volume} {38}},\ \bibinfo {pages} {095004}
  (\bibinfo {year} {2021})},\ \Eprint {http://arxiv.org/abs/2012.11512}
  {arXiv:2012.11512 [gr-qc]} \BibitemShut {NoStop}%
\bibitem [{\citenamefont {Hooper}\ \emph {et~al.}(2012)\citenamefont {Hooper},
  \citenamefont {Chung}, \citenamefont {Luan}, \citenamefont {Blair},
  \citenamefont {Chen},\ and\ \citenamefont {Wen}}]{Hooper:2011rb}%
  \BibitemOpen
  \bibfield  {author} {\bibinfo {author} {\bibfnamefont {Shaun}\ \bibnamefont
  {Hooper}}, \bibinfo {author} {\bibfnamefont {Shin~Kee}\ \bibnamefont
  {Chung}}, \bibinfo {author} {\bibfnamefont {Jing}\ \bibnamefont {Luan}},
  \bibinfo {author} {\bibfnamefont {David}\ \bibnamefont {Blair}}, \bibinfo
  {author} {\bibfnamefont {Yanbei}\ \bibnamefont {Chen}}, \ and\ \bibinfo
  {author} {\bibfnamefont {Linqing}\ \bibnamefont {Wen}},\ }\bibfield  {title}
  {\enquote {\bibinfo {title} {{Summed Parallel Infinite Impulse Response
  (SPIIR) Filters For Low-Latency Gravitational Wave Detection}},}\ }\href
  {\doibase 10.1103/PhysRevD.86.024012} {\bibfield  {journal} {\bibinfo
  {journal} {Phys. Rev. D}\ }\textbf {\bibinfo {volume} {86}},\ \bibinfo
  {pages} {024012} (\bibinfo {year} {2012})},\ \Eprint
  {http://arxiv.org/abs/1108.3186} {arXiv:1108.3186 [gr-qc]} \BibitemShut
  {NoStop}%
\bibitem [{\citenamefont {Pratten}\ \emph {et~al.}(2021)\citenamefont {Pratten}
  \emph {et~al.}}]{Pratten:2020ceb}%
  \BibitemOpen
  \bibfield  {author} {\bibinfo {author} {\bibfnamefont {Geraint}\ \bibnamefont
  {Pratten}} \emph {et~al.},\ }\bibfield  {title} {\enquote {\bibinfo {title}
  {{Computationally efficient models for the dominant and subdominant harmonic
  modes of precessing binary black holes}},}\ }\href {\doibase
  10.1103/PhysRevD.103.104056} {\bibfield  {journal} {\bibinfo  {journal}
  {Phys. Rev. D}\ }\textbf {\bibinfo {volume} {103}},\ \bibinfo {pages}
  {104056} (\bibinfo {year} {2021})},\ \Eprint
  {http://arxiv.org/abs/2004.06503} {arXiv:2004.06503 [gr-qc]} \BibitemShut
  {NoStop}%
\bibitem [{\citenamefont {Ossokine}\ \emph {et~al.}(2020)\citenamefont
  {Ossokine} \emph {et~al.}}]{Ossokine:2020kjp}%
  \BibitemOpen
  \bibfield  {author} {\bibinfo {author} {\bibfnamefont {Serguei}\ \bibnamefont
  {Ossokine}} \emph {et~al.},\ }\bibfield  {title} {\enquote {\bibinfo {title}
  {{Multipolar Effective-One-Body Waveforms for Precessing Binary Black Holes:
  Construction and Validation}},}\ }\href {\doibase
  10.1103/PhysRevD.102.044055} {\bibfield  {journal} {\bibinfo  {journal}
  {Phys. Rev. D}\ }\textbf {\bibinfo {volume} {102}},\ \bibinfo {pages}
  {044055} (\bibinfo {year} {2020})},\ \Eprint
  {http://arxiv.org/abs/2004.09442} {arXiv:2004.09442 [gr-qc]} \BibitemShut
  {NoStop}%
\bibitem [{\citenamefont {Harry}\ \emph {et~al.}(2016)\citenamefont {Harry},
  \citenamefont {Privitera}, \citenamefont {Boh\'e},\ and\ \citenamefont
  {Buonanno}}]{Harry:2016ijz}%
  \BibitemOpen
  \bibfield  {author} {\bibinfo {author} {\bibfnamefont {Ian}\ \bibnamefont
  {Harry}}, \bibinfo {author} {\bibfnamefont {Stephen}\ \bibnamefont
  {Privitera}}, \bibinfo {author} {\bibfnamefont {Alejandro}\ \bibnamefont
  {Boh\'e}}, \ and\ \bibinfo {author} {\bibfnamefont {Alessandra}\ \bibnamefont
  {Buonanno}},\ }\bibfield  {title} {\enquote {\bibinfo {title} {{Searching for
  Gravitational Waves from Compact Binaries with Precessing Spins}},}\ }\href
  {\doibase 10.1103/PhysRevD.94.024012} {\bibfield  {journal} {\bibinfo
  {journal} {Phys. Rev. D}\ }\textbf {\bibinfo {volume} {94}},\ \bibinfo
  {pages} {024012} (\bibinfo {year} {2016})},\ \Eprint
  {http://arxiv.org/abs/1603.02444} {arXiv:1603.02444 [gr-qc]} \BibitemShut
  {NoStop}%
\bibitem [{\citenamefont {Harry}\ \emph {et~al.}(2018)\citenamefont {Harry},
  \citenamefont {Calder\'on~Bustillo},\ and\ \citenamefont
  {Nitz}}]{Harry:2017weg}%
  \BibitemOpen
  \bibfield  {author} {\bibinfo {author} {\bibfnamefont {Ian}\ \bibnamefont
  {Harry}}, \bibinfo {author} {\bibfnamefont {Juan}\ \bibnamefont
  {Calder\'on~Bustillo}}, \ and\ \bibinfo {author} {\bibfnamefont {Alex}\
  \bibnamefont {Nitz}},\ }\bibfield  {title} {\enquote {\bibinfo {title}
  {{Searching for the full symphony of black hole binary mergers}},}\ }\href
  {\doibase 10.1103/PhysRevD.97.023004} {\bibfield  {journal} {\bibinfo
  {journal} {Phys. Rev. D}\ }\textbf {\bibinfo {volume} {97}},\ \bibinfo
  {pages} {023004} (\bibinfo {year} {2018})},\ \Eprint
  {http://arxiv.org/abs/1709.09181} {arXiv:1709.09181 [gr-qc]} \BibitemShut
  {NoStop}%
\bibitem [{\citenamefont {Chandra}\ \emph {et~al.}(2022)\citenamefont
  {Chandra}, \citenamefont {Calder\'on~Bustillo}, \citenamefont {Pai},\ and\
  \citenamefont {Harry}}]{Chandra:2022ixv}%
  \BibitemOpen
  \bibfield  {author} {\bibinfo {author} {\bibfnamefont {Koustav}\ \bibnamefont
  {Chandra}}, \bibinfo {author} {\bibfnamefont {Juan}\ \bibnamefont
  {Calder\'on~Bustillo}}, \bibinfo {author} {\bibfnamefont {Archana}\
  \bibnamefont {Pai}}, \ and\ \bibinfo {author} {\bibfnamefont {Ian}\
  \bibnamefont {Harry}},\ }\bibfield  {title} {\enquote {\bibinfo {title}
  {{First gravitational-wave search for intermediate-mass black hole mergers
  with higher order harmonics}},}\ }\href@noop {} {\  (\bibinfo {year}
  {2022})},\ \Eprint {http://arxiv.org/abs/2207.01654} {arXiv:2207.01654
  [gr-qc]} \BibitemShut {NoStop}%
\bibitem [{\citenamefont {Calder\'on~Bustillo}\ \emph
  {et~al.}(2017)\citenamefont {Calder\'on~Bustillo}, \citenamefont {Laguna},\
  and\ \citenamefont {Shoemaker}}]{CalderonBustillo:2016rlt}%
  \BibitemOpen
  \bibfield  {author} {\bibinfo {author} {\bibfnamefont {Juan}\ \bibnamefont
  {Calder\'on~Bustillo}}, \bibinfo {author} {\bibfnamefont {Pablo}\
  \bibnamefont {Laguna}}, \ and\ \bibinfo {author} {\bibfnamefont {Deirdre}\
  \bibnamefont {Shoemaker}},\ }\bibfield  {title} {\enquote {\bibinfo {title}
  {{Detectability of gravitational waves from binary black holes: Impact of
  precession and higher modes}},}\ }\href {\doibase 10.1103/PhysRevD.95.104038}
  {\bibfield  {journal} {\bibinfo  {journal} {Phys. Rev. D}\ }\textbf {\bibinfo
  {volume} {95}},\ \bibinfo {pages} {104038} (\bibinfo {year} {2017})},\
  \Eprint {http://arxiv.org/abs/1612.02340} {arXiv:1612.02340 [gr-qc]}
  \BibitemShut {NoStop}%
\bibitem [{\citenamefont {Blackman}\ \emph {et~al.}(2017)\citenamefont
  {Blackman}, \citenamefont {Field}, \citenamefont {Scheel}, \citenamefont
  {Galley}, \citenamefont {Ott}, \citenamefont {Boyle}, \citenamefont {Kidder},
  \citenamefont {Pfeiffer},\ and\ \citenamefont
  {Szil\'agyi}}]{Blackman:2017pcm}%
  \BibitemOpen
  \bibfield  {author} {\bibinfo {author} {\bibfnamefont {Jonathan}\
  \bibnamefont {Blackman}}, \bibinfo {author} {\bibfnamefont {Scott~E.}\
  \bibnamefont {Field}}, \bibinfo {author} {\bibfnamefont {Mark~A.}\
  \bibnamefont {Scheel}}, \bibinfo {author} {\bibfnamefont {Chad~R.}\
  \bibnamefont {Galley}}, \bibinfo {author} {\bibfnamefont {Christian~D.}\
  \bibnamefont {Ott}}, \bibinfo {author} {\bibfnamefont {Michael}\ \bibnamefont
  {Boyle}}, \bibinfo {author} {\bibfnamefont {Lawrence~E.}\ \bibnamefont
  {Kidder}}, \bibinfo {author} {\bibfnamefont {Harald~P.}\ \bibnamefont
  {Pfeiffer}}, \ and\ \bibinfo {author} {\bibfnamefont {B\'ela}\ \bibnamefont
  {Szil\'agyi}},\ }\bibfield  {title} {\enquote {\bibinfo {title} {{Numerical
  relativity waveform surrogate model for generically precessing binary black
  hole mergers}},}\ }\href {\doibase 10.1103/PhysRevD.96.024058} {\bibfield
  {journal} {\bibinfo  {journal} {Phys. Rev. D}\ }\textbf {\bibinfo {volume}
  {96}},\ \bibinfo {pages} {024058} (\bibinfo {year} {2017})},\ \Eprint
  {http://arxiv.org/abs/1705.07089} {arXiv:1705.07089 [gr-qc]} \BibitemShut
  {NoStop}%
\bibitem [{\citenamefont {Mills}\ and\ \citenamefont
  {Fairhurst}(2021)}]{Mills:2020thr}%
  \BibitemOpen
  \bibfield  {author} {\bibinfo {author} {\bibfnamefont {Cameron}\ \bibnamefont
  {Mills}}\ and\ \bibinfo {author} {\bibfnamefont {Stephen}\ \bibnamefont
  {Fairhurst}},\ }\bibfield  {title} {\enquote {\bibinfo {title} {{Measuring
  gravitational-wave higher-order multipoles}},}\ }\href {\doibase
  10.1103/PhysRevD.103.024042} {\bibfield  {journal} {\bibinfo  {journal}
  {Phys. Rev. D}\ }\textbf {\bibinfo {volume} {103}},\ \bibinfo {pages}
  {024042} (\bibinfo {year} {2021})},\ \Eprint
  {http://arxiv.org/abs/2007.04313} {arXiv:2007.04313 [gr-qc]} \BibitemShut
  {NoStop}%
\bibitem [{\citenamefont {Apostolatos}\ \emph {et~al.}(1994)\citenamefont
  {Apostolatos}, \citenamefont {Cutler}, \citenamefont {Sussman},\ and\
  \citenamefont {Thorne}}]{Apostolatos:1994mx}%
  \BibitemOpen
  \bibfield  {author} {\bibinfo {author} {\bibfnamefont {Theocharis~A.}\
  \bibnamefont {Apostolatos}}, \bibinfo {author} {\bibfnamefont {Curt}\
  \bibnamefont {Cutler}}, \bibinfo {author} {\bibfnamefont {Gerald~J.}\
  \bibnamefont {Sussman}}, \ and\ \bibinfo {author} {\bibfnamefont {Kip~S.}\
  \bibnamefont {Thorne}},\ }\bibfield  {title} {\enquote {\bibinfo {title}
  {{Spin induced orbital precession and its modulation of the gravitational
  wave forms from merging binaries}},}\ }\href {\doibase
  10.1103/PhysRevD.49.6274} {\bibfield  {journal} {\bibinfo  {journal} {Phys.
  Rev. D}\ }\textbf {\bibinfo {volume} {49}},\ \bibinfo {pages} {6274--6297}
  (\bibinfo {year} {1994})}\BibitemShut {NoStop}%
\bibitem [{\citenamefont {Ajith}\ \emph {et~al.}(2011)\citenamefont {Ajith}
  \emph {et~al.}}]{Ajith:2009bn}%
  \BibitemOpen
  \bibfield  {author} {\bibinfo {author} {\bibfnamefont {P.}~\bibnamefont
  {Ajith}} \emph {et~al.},\ }\bibfield  {title} {\enquote {\bibinfo {title}
  {{Inspiral-merger-ringdown waveforms for black-hole binaries with
  non-precessing spins}},}\ }\href {\doibase 10.1103/PhysRevLett.106.241101}
  {\bibfield  {journal} {\bibinfo  {journal} {Phys. Rev. Lett.}\ }\textbf
  {\bibinfo {volume} {106}},\ \bibinfo {pages} {241101} (\bibinfo {year}
  {2011})},\ \Eprint {http://arxiv.org/abs/0909.2867} {arXiv:0909.2867 [gr-qc]}
  \BibitemShut {NoStop}%
\bibitem [{\citenamefont {Schmidt}\ \emph {et~al.}(2015)\citenamefont
  {Schmidt}, \citenamefont {Ohme},\ and\ \citenamefont
  {Hannam}}]{Schmidt:2014iyl}%
  \BibitemOpen
  \bibfield  {author} {\bibinfo {author} {\bibfnamefont {Patricia}\
  \bibnamefont {Schmidt}}, \bibinfo {author} {\bibfnamefont {Frank}\
  \bibnamefont {Ohme}}, \ and\ \bibinfo {author} {\bibfnamefont {Mark}\
  \bibnamefont {Hannam}},\ }\bibfield  {title} {\enquote {\bibinfo {title}
  {{Towards models of gravitational waveforms from generic binaries II:
  Modelling precession effects with a single effective precession
  parameter}},}\ }\href {\doibase 10.1103/PhysRevD.91.024043} {\bibfield
  {journal} {\bibinfo  {journal} {Phys. Rev. D}\ }\textbf {\bibinfo {volume}
  {91}},\ \bibinfo {pages} {024043} (\bibinfo {year} {2015})},\ \Eprint
  {http://arxiv.org/abs/1408.1810} {arXiv:1408.1810 [gr-qc]} \BibitemShut
  {NoStop}%
\bibitem [{\citenamefont {Estell\'es}\ \emph {et~al.}(2022)\citenamefont
  {Estell\'es} \emph {et~al.}}]{Estelles:2021jnz}%
  \BibitemOpen
  \bibfield  {author} {\bibinfo {author} {\bibfnamefont {H\'ector}\
  \bibnamefont {Estell\'es}} \emph {et~al.},\ }\bibfield  {title} {\enquote
  {\bibinfo {title} {{A Detailed Analysis of GW190521 with Phenomenological
  Waveform Models}},}\ }\href {\doibase 10.3847/1538-4357/ac33a0} {\bibfield
  {journal} {\bibinfo  {journal} {Astrophys. J.}\ }\textbf {\bibinfo {volume}
  {924}},\ \bibinfo {pages} {79} (\bibinfo {year} {2022})},\ \Eprint
  {http://arxiv.org/abs/2105.06360} {arXiv:2105.06360 [gr-qc]} \BibitemShut
  {NoStop}%
\bibitem [{\citenamefont {Tiwari}(2022)}]{Tiwari:2021yvr}%
  \BibitemOpen
  \bibfield  {author} {\bibinfo {author} {\bibfnamefont {Vaibhav}\ \bibnamefont
  {Tiwari}},\ }\bibfield  {title} {\enquote {\bibinfo {title} {{Exploring
  Features in the Binary Black Hole Population}},}\ }\href {\doibase
  10.3847/1538-4357/ac589a} {\bibfield  {journal} {\bibinfo  {journal}
  {Astrophys. J.}\ }\textbf {\bibinfo {volume} {928}},\ \bibinfo {pages} {155}
  (\bibinfo {year} {2022})},\ \Eprint {http://arxiv.org/abs/2111.13991}
  {arXiv:2111.13991 [astro-ph.HE]} \BibitemShut {NoStop}%
\bibitem [{\citenamefont {Zhu}\ \emph {et~al.}(2022)\citenamefont {Zhu},
  \citenamefont {Wu}, \citenamefont {Qin}, \citenamefont {Zhang}, \citenamefont
  {Gao},\ and\ \citenamefont {Cao}}]{Zhu:2021jbw}%
  \BibitemOpen
  \bibfield  {author} {\bibinfo {author} {\bibfnamefont {Jin-Ping}\
  \bibnamefont {Zhu}}, \bibinfo {author} {\bibfnamefont {Shichao}\ \bibnamefont
  {Wu}}, \bibinfo {author} {\bibfnamefont {Ying}\ \bibnamefont {Qin}}, \bibinfo
  {author} {\bibfnamefont {Bing}\ \bibnamefont {Zhang}}, \bibinfo {author}
  {\bibfnamefont {He}~\bibnamefont {Gao}}, \ and\ \bibinfo {author}
  {\bibfnamefont {Zhoujian}\ \bibnamefont {Cao}},\ }\bibfield  {title}
  {\enquote {\bibinfo {title} {{Population Properties of Gravitational-wave
  Neutron Star\textendash{}Black Hole Mergers}},}\ }\href {\doibase
  10.3847/1538-4357/ac540c} {\bibfield  {journal} {\bibinfo  {journal}
  {Astrophys. J.}\ }\textbf {\bibinfo {volume} {928}},\ \bibinfo {pages} {167}
  (\bibinfo {year} {2022})},\ \Eprint {http://arxiv.org/abs/2112.02605}
  {arXiv:2112.02605 [astro-ph.HE]} \BibitemShut {NoStop}%
\bibitem [{\citenamefont {Nitz}\ \emph {et~al.}(2019)\citenamefont {Nitz},
  \citenamefont {Capano}, \citenamefont {Nielsen}, \citenamefont {Reyes},
  \citenamefont {White}, \citenamefont {Brown},\ and\ \citenamefont
  {Krishnan}}]{Nitz:2018imz}%
  \BibitemOpen
  \bibfield  {author} {\bibinfo {author} {\bibfnamefont {Alexander~H.}\
  \bibnamefont {Nitz}}, \bibinfo {author} {\bibfnamefont {Collin}\ \bibnamefont
  {Capano}}, \bibinfo {author} {\bibfnamefont {Alex~B.}\ \bibnamefont
  {Nielsen}}, \bibinfo {author} {\bibfnamefont {Steven}\ \bibnamefont {Reyes}},
  \bibinfo {author} {\bibfnamefont {Rebecca}\ \bibnamefont {White}}, \bibinfo
  {author} {\bibfnamefont {Duncan~A.}\ \bibnamefont {Brown}}, \ and\ \bibinfo
  {author} {\bibfnamefont {Badri}\ \bibnamefont {Krishnan}},\ }\bibfield
  {title} {\enquote {\bibinfo {title} {{1-OGC: The first open
  gravitational-wave catalog of binary mergers from analysis of public Advanced
  LIGO data}},}\ }\href {\doibase 10.3847/1538-4357/ab0108} {\bibfield
  {journal} {\bibinfo  {journal} {Astrophys. J.}\ }\textbf {\bibinfo {volume}
  {872}},\ \bibinfo {pages} {195} (\bibinfo {year} {2019})},\ \Eprint
  {http://arxiv.org/abs/1811.01921} {arXiv:1811.01921 [gr-qc]} \BibitemShut
  {NoStop}%
\bibitem [{\citenamefont {Abbott}\ \emph
  {et~al.}(2018{\natexlab{b}})\citenamefont {Abbott} \emph
  {et~al.}}]{LIGOScientific:2017tza}%
  \BibitemOpen
  \bibfield  {author} {\bibinfo {author} {\bibfnamefont {B~P}\ \bibnamefont
  {Abbott}} \emph {et~al.} (\bibinfo {collaboration} {LIGO Scientific,
  Virgo}),\ }\bibfield  {title} {\enquote {\bibinfo {title} {{Effects of data
  quality vetoes on a search for compact binary coalescences in Advanced
  LIGO\textquoteright{}s first observing run}},}\ }\href {\doibase
  10.1088/1361-6382/aaaafa} {\bibfield  {journal} {\bibinfo  {journal} {Class.
  Quant. Grav.}\ }\textbf {\bibinfo {volume} {35}},\ \bibinfo {pages} {065010}
  (\bibinfo {year} {2018}{\natexlab{b}})},\ \Eprint
  {http://arxiv.org/abs/1710.02185} {arXiv:1710.02185 [gr-qc]} \BibitemShut
  {NoStop}%
\bibitem [{\citenamefont {Allen}(2005)}]{Allen:2004gu}%
  \BibitemOpen
  \bibfield  {author} {\bibinfo {author} {\bibfnamefont {Bruce}\ \bibnamefont
  {Allen}},\ }\bibfield  {title} {\enquote {\bibinfo {title} {{${\chi}^{2}$
  time-frequency discriminator for gravitational wave detection}},}\ }\href
  {\doibase 10.1103/PhysRevD.71.062001} {\bibfield  {journal} {\bibinfo
  {journal} {Phys. Rev. D}\ }\textbf {\bibinfo {volume} {71}},\ \bibinfo
  {pages} {062001} (\bibinfo {year} {2005})},\ \Eprint
  {http://arxiv.org/abs/gr-qc/0405045} {arXiv:gr-qc/0405045} \BibitemShut
  {NoStop}%
\bibitem [{\citenamefont {Abbott}\ \emph
  {et~al.}(2016{\natexlab{b}})\citenamefont {Abbott} \emph
  {et~al.}}]{LIGOScientific:2016gtq}%
  \BibitemOpen
  \bibfield  {author} {\bibinfo {author} {\bibfnamefont {B.~P.}\ \bibnamefont
  {Abbott}} \emph {et~al.} (\bibinfo {collaboration} {LIGO Scientific,
  Virgo}),\ }\bibfield  {title} {\enquote {\bibinfo {title} {{Characterization
  of transient noise in Advanced LIGO relevant to gravitational wave signal
  GW150914}},}\ }\href {\doibase 10.1088/0264-9381/33/13/134001} {\bibfield
  {journal} {\bibinfo  {journal} {Class. Quant. Grav.}\ }\textbf {\bibinfo
  {volume} {33}},\ \bibinfo {pages} {134001} (\bibinfo {year}
  {2016}{\natexlab{b}})},\ \Eprint {http://arxiv.org/abs/1602.03844}
  {arXiv:1602.03844 [gr-qc]} \BibitemShut {NoStop}%
\bibitem [{\citenamefont {Davis}\ \emph {et~al.}(2021)\citenamefont {Davis}
  \emph {et~al.}}]{LIGO:2021ppb}%
  \BibitemOpen
  \bibfield  {author} {\bibinfo {author} {\bibfnamefont {Derek}\ \bibnamefont
  {Davis}} \emph {et~al.} (\bibinfo {collaboration} {LIGO}),\ }\bibfield
  {title} {\enquote {\bibinfo {title} {{LIGO detector characterization in the
  second and third observing runs}},}\ }\href {\doibase
  10.1088/1361-6382/abfd85} {\bibfield  {journal} {\bibinfo  {journal} {Class.
  Quant. Grav.}\ }\textbf {\bibinfo {volume} {38}},\ \bibinfo {pages} {135014}
  (\bibinfo {year} {2021})},\ \Eprint {http://arxiv.org/abs/2101.11673}
  {arXiv:2101.11673 [astro-ph.IM]} \BibitemShut {NoStop}%
\bibitem [{\citenamefont {Abbott}\ \emph
  {et~al.}(2019{\natexlab{b}})\citenamefont {Abbott} \emph
  {et~al.}}]{LIGOScientific:2018mvr}%
  \BibitemOpen
  \bibfield  {author} {\bibinfo {author} {\bibfnamefont {B.~P.}\ \bibnamefont
  {Abbott}} \emph {et~al.} (\bibinfo {collaboration} {LIGO Scientific,
  Virgo}),\ }\bibfield  {title} {\enquote {\bibinfo {title} {{GWTC-1: A
  Gravitational-Wave Transient Catalog of Compact Binary Mergers Observed by
  LIGO and Virgo during the First and Second Observing Runs}},}\ }\href
  {\doibase 10.1103/PhysRevX.9.031040} {\bibfield  {journal} {\bibinfo
  {journal} {Phys. Rev. X}\ }\textbf {\bibinfo {volume} {9}},\ \bibinfo {pages}
  {031040} (\bibinfo {year} {2019}{\natexlab{b}})},\ \Eprint
  {http://arxiv.org/abs/1811.12907} {arXiv:1811.12907 [astro-ph.HE]}
  \BibitemShut {NoStop}%
\bibitem [{\citenamefont {Allen}\ \emph {et~al.}(2012)\citenamefont {Allen},
  \citenamefont {Anderson}, \citenamefont {Brady}, \citenamefont {Brown},\ and\
  \citenamefont {Creighton}}]{Allen:2005fk}%
  \BibitemOpen
  \bibfield  {author} {\bibinfo {author} {\bibfnamefont {Bruce}\ \bibnamefont
  {Allen}}, \bibinfo {author} {\bibfnamefont {Warren~G.}\ \bibnamefont
  {Anderson}}, \bibinfo {author} {\bibfnamefont {Patrick~R.}\ \bibnamefont
  {Brady}}, \bibinfo {author} {\bibfnamefont {Duncan~A.}\ \bibnamefont
  {Brown}}, \ and\ \bibinfo {author} {\bibfnamefont {Jolien D.~E.}\
  \bibnamefont {Creighton}},\ }\bibfield  {title} {\enquote {\bibinfo {title}
  {{FINDCHIRP: An Algorithm for detection of gravitational waves from
  inspiraling compact binaries}},}\ }\href {\doibase
  10.1103/PhysRevD.85.122006} {\bibfield  {journal} {\bibinfo  {journal} {Phys.
  Rev. D}\ }\textbf {\bibinfo {volume} {85}},\ \bibinfo {pages} {122006}
  (\bibinfo {year} {2012})},\ \Eprint {http://arxiv.org/abs/gr-qc/0509116}
  {arXiv:gr-qc/0509116} \BibitemShut {NoStop}%
\bibitem [{\citenamefont {Owen}\ and\ \citenamefont
  {Sathyaprakash}(1999)}]{Owen:1998dk}%
  \BibitemOpen
  \bibfield  {author} {\bibinfo {author} {\bibfnamefont {Benjamin~J.}\
  \bibnamefont {Owen}}\ and\ \bibinfo {author} {\bibfnamefont {B.~S.}\
  \bibnamefont {Sathyaprakash}},\ }\bibfield  {title} {\enquote {\bibinfo
  {title} {{Matched filtering of gravitational waves from inspiraling compact
  binaries: Computational cost and template placement}},}\ }\href {\doibase
  10.1103/PhysRevD.60.022002} {\bibfield  {journal} {\bibinfo  {journal} {Phys.
  Rev. D}\ }\textbf {\bibinfo {volume} {60}},\ \bibinfo {pages} {022002}
  (\bibinfo {year} {1999})},\ \Eprint {http://arxiv.org/abs/gr-qc/9808076}
  {arXiv:gr-qc/9808076} \BibitemShut {NoStop}%
\bibitem [{\citenamefont {Babak}\ \emph {et~al.}(2006)\citenamefont {Babak},
  \citenamefont {Balasubramanian}, \citenamefont {Churches}, \citenamefont
  {Cokelaer},\ and\ \citenamefont {Sathyaprakash}}]{Babak:2006ty}%
  \BibitemOpen
  \bibfield  {author} {\bibinfo {author} {\bibfnamefont {S.}~\bibnamefont
  {Babak}}, \bibinfo {author} {\bibfnamefont {R.}~\bibnamefont
  {Balasubramanian}}, \bibinfo {author} {\bibfnamefont {D.}~\bibnamefont
  {Churches}}, \bibinfo {author} {\bibfnamefont {T.}~\bibnamefont {Cokelaer}},
  \ and\ \bibinfo {author} {\bibfnamefont {B.~S.}\ \bibnamefont
  {Sathyaprakash}},\ }\bibfield  {title} {\enquote {\bibinfo {title} {{A
  Template bank to search for gravitational waves from inspiralling compact
  binaries. I. Physical models}},}\ }\href {\doibase
  10.1088/0264-9381/23/18/002} {\bibfield  {journal} {\bibinfo  {journal}
  {Class. Quant. Grav.}\ }\textbf {\bibinfo {volume} {23}},\ \bibinfo {pages}
  {5477--5504} (\bibinfo {year} {2006})},\ \Eprint
  {http://arxiv.org/abs/gr-qc/0604037} {arXiv:gr-qc/0604037} \BibitemShut
  {NoStop}%
\bibitem [{\citenamefont {Harry}\ \emph {et~al.}(2009)\citenamefont {Harry},
  \citenamefont {Allen},\ and\ \citenamefont {Sathyaprakash}}]{Harry:2009ea}%
  \BibitemOpen
  \bibfield  {author} {\bibinfo {author} {\bibfnamefont {Ian~W.}\ \bibnamefont
  {Harry}}, \bibinfo {author} {\bibfnamefont {Bruce}\ \bibnamefont {Allen}}, \
  and\ \bibinfo {author} {\bibfnamefont {B.~S.}\ \bibnamefont
  {Sathyaprakash}},\ }\bibfield  {title} {\enquote {\bibinfo {title} {{A
  Stochastic template placement algorithm for gravitational wave data
  analysis}},}\ }\href {\doibase 10.1103/PhysRevD.80.104014} {\bibfield
  {journal} {\bibinfo  {journal} {Phys. Rev. D}\ }\textbf {\bibinfo {volume}
  {80}},\ \bibinfo {pages} {104014} (\bibinfo {year} {2009})},\ \Eprint
  {http://arxiv.org/abs/0908.2090} {arXiv:0908.2090 [gr-qc]} \BibitemShut
  {NoStop}%
\bibitem [{\citenamefont {Babak}(2008)}]{Babak:2008rb}%
  \BibitemOpen
  \bibfield  {author} {\bibinfo {author} {\bibfnamefont {Stanislav}\
  \bibnamefont {Babak}},\ }\bibfield  {title} {\enquote {\bibinfo {title}
  {{Building a stochastic template bank for detecting massive black hole
  binaries}},}\ }\href {\doibase 10.1088/0264-9381/25/19/195011} {\bibfield
  {journal} {\bibinfo  {journal} {Class. Quant. Grav.}\ }\textbf {\bibinfo
  {volume} {25}},\ \bibinfo {pages} {195011} (\bibinfo {year} {2008})},\
  \Eprint {http://arxiv.org/abs/0801.4070} {arXiv:0801.4070 [gr-qc]}
  \BibitemShut {NoStop}%
\bibitem [{\citenamefont {Roy}\ \emph {et~al.}(2019)\citenamefont {Roy},
  \citenamefont {Sengupta},\ and\ \citenamefont {Ajith}}]{Roy:2017oul}%
  \BibitemOpen
  \bibfield  {author} {\bibinfo {author} {\bibfnamefont {Soumen}\ \bibnamefont
  {Roy}}, \bibinfo {author} {\bibfnamefont {Anand~S.}\ \bibnamefont
  {Sengupta}}, \ and\ \bibinfo {author} {\bibfnamefont {Parameswaran}\
  \bibnamefont {Ajith}},\ }\bibfield  {title} {\enquote {\bibinfo {title}
  {{Effectual template banks for upcoming compact binary searches in
  Advanced-LIGO and Virgo data}},}\ }\href {\doibase
  10.1103/PhysRevD.99.024048} {\bibfield  {journal} {\bibinfo  {journal} {Phys.
  Rev. D}\ }\textbf {\bibinfo {volume} {99}},\ \bibinfo {pages} {024048}
  (\bibinfo {year} {2019})},\ \Eprint {http://arxiv.org/abs/1711.08743}
  {arXiv:1711.08743 [gr-qc]} \BibitemShut {NoStop}%
\bibitem [{\citenamefont {Capano}\ \emph {et~al.}(2016)\citenamefont {Capano},
  \citenamefont {Harry}, \citenamefont {Privitera},\ and\ \citenamefont
  {Buonanno}}]{Capano:2016dsf}%
  \BibitemOpen
  \bibfield  {author} {\bibinfo {author} {\bibfnamefont {Collin}\ \bibnamefont
  {Capano}}, \bibinfo {author} {\bibfnamefont {Ian}\ \bibnamefont {Harry}},
  \bibinfo {author} {\bibfnamefont {Stephen}\ \bibnamefont {Privitera}}, \ and\
  \bibinfo {author} {\bibfnamefont {Alessandra}\ \bibnamefont {Buonanno}},\
  }\bibfield  {title} {\enquote {\bibinfo {title} {{Implementing a search for
  gravitational waves from binary black holes with nonprecessing spin}},}\
  }\href {\doibase 10.1103/PhysRevD.93.124007} {\bibfield  {journal} {\bibinfo
  {journal} {Phys. Rev. D}\ }\textbf {\bibinfo {volume} {93}},\ \bibinfo
  {pages} {124007} (\bibinfo {year} {2016})},\ \Eprint
  {http://arxiv.org/abs/1602.03509} {arXiv:1602.03509 [gr-qc]} \BibitemShut
  {NoStop}%
\bibitem [{\citenamefont {Lenon}\ \emph {et~al.}(2021)\citenamefont {Lenon},
  \citenamefont {Brown},\ and\ \citenamefont {Nitz}}]{Lenon:2021zac}%
  \BibitemOpen
  \bibfield  {author} {\bibinfo {author} {\bibfnamefont {Amber~K.}\
  \bibnamefont {Lenon}}, \bibinfo {author} {\bibfnamefont {Duncan~A.}\
  \bibnamefont {Brown}}, \ and\ \bibinfo {author} {\bibfnamefont
  {Alexander~H.}\ \bibnamefont {Nitz}},\ }\bibfield  {title} {\enquote
  {\bibinfo {title} {{Eccentric binary neutron star search prospects for Cosmic
  Explorer}},}\ }\href {\doibase 10.1103/PhysRevD.104.063011} {\bibfield
  {journal} {\bibinfo  {journal} {Phys. Rev. D}\ }\textbf {\bibinfo {volume}
  {104}},\ \bibinfo {pages} {063011} (\bibinfo {year} {2021})},\ \Eprint
  {http://arxiv.org/abs/2103.14088} {arXiv:2103.14088 [astro-ph.HE]}
  \BibitemShut {NoStop}%
\bibitem [{\citenamefont {Buonanno}\ \emph {et~al.}(2003)\citenamefont
  {Buonanno}, \citenamefont {Chen},\ and\ \citenamefont
  {Vallisneri}}]{Buonanno:2002fy}%
  \BibitemOpen
  \bibfield  {author} {\bibinfo {author} {\bibfnamefont {Alessandra}\
  \bibnamefont {Buonanno}}, \bibinfo {author} {\bibfnamefont {Yan-bei}\
  \bibnamefont {Chen}}, \ and\ \bibinfo {author} {\bibfnamefont {Michele}\
  \bibnamefont {Vallisneri}},\ }\bibfield  {title} {\enquote {\bibinfo {title}
  {{Detecting gravitational waves from precessing binaries of spinning compact
  objects: Adiabatic limit}},}\ }\href {\doibase 10.1103/PhysRevD.67.104025}
  {\bibfield  {journal} {\bibinfo  {journal} {Phys. Rev. D}\ }\textbf {\bibinfo
  {volume} {67}},\ \bibinfo {pages} {104025} (\bibinfo {year} {2003})},\
  \bibinfo {note} {[Erratum: Phys.Rev.D 74, 029904 (2006)]},\ \Eprint
  {http://arxiv.org/abs/gr-qc/0211087} {arXiv:gr-qc/0211087} \BibitemShut
  {NoStop}%
\bibitem [{\citenamefont {Husa}\ \emph {et~al.}(2016)\citenamefont {Husa},
  \citenamefont {Khan}, \citenamefont {Hannam}, \citenamefont {P\"urrer},
  \citenamefont {Ohme}, \citenamefont {Jim\'enez~Forteza},\ and\ \citenamefont
  {Boh\'e}}]{Husa:2015iqa}%
  \BibitemOpen
  \bibfield  {author} {\bibinfo {author} {\bibfnamefont {Sascha}\ \bibnamefont
  {Husa}}, \bibinfo {author} {\bibfnamefont {Sebastian}\ \bibnamefont {Khan}},
  \bibinfo {author} {\bibfnamefont {Mark}\ \bibnamefont {Hannam}}, \bibinfo
  {author} {\bibfnamefont {Michael}\ \bibnamefont {P\"urrer}}, \bibinfo
  {author} {\bibfnamefont {Frank}\ \bibnamefont {Ohme}}, \bibinfo {author}
  {\bibfnamefont {Xisco}\ \bibnamefont {Jim\'enez~Forteza}}, \ and\ \bibinfo
  {author} {\bibfnamefont {Alejandro}\ \bibnamefont {Boh\'e}},\ }\bibfield
  {title} {\enquote {\bibinfo {title} {{Frequency-domain gravitational waves
  from nonprecessing black-hole binaries. I. New numerical waveforms and
  anatomy of the signal}},}\ }\href {\doibase 10.1103/PhysRevD.93.044006}
  {\bibfield  {journal} {\bibinfo  {journal} {Phys. Rev. D}\ }\textbf {\bibinfo
  {volume} {93}},\ \bibinfo {pages} {044006} (\bibinfo {year} {2016})},\
  \Eprint {http://arxiv.org/abs/1508.07250} {arXiv:1508.07250 [gr-qc]}
  \BibitemShut {NoStop}%
\bibitem [{\citenamefont {Khan}\ \emph {et~al.}(2016)\citenamefont {Khan},
  \citenamefont {Husa}, \citenamefont {Hannam}, \citenamefont {Ohme},
  \citenamefont {P\"urrer}, \citenamefont {Jim\'enez~Forteza},\ and\
  \citenamefont {Boh\'e}}]{Khan:2015jqa}%
  \BibitemOpen
  \bibfield  {author} {\bibinfo {author} {\bibfnamefont {Sebastian}\
  \bibnamefont {Khan}}, \bibinfo {author} {\bibfnamefont {Sascha}\ \bibnamefont
  {Husa}}, \bibinfo {author} {\bibfnamefont {Mark}\ \bibnamefont {Hannam}},
  \bibinfo {author} {\bibfnamefont {Frank}\ \bibnamefont {Ohme}}, \bibinfo
  {author} {\bibfnamefont {Michael}\ \bibnamefont {P\"urrer}}, \bibinfo
  {author} {\bibfnamefont {Xisco}\ \bibnamefont {Jim\'enez~Forteza}}, \ and\
  \bibinfo {author} {\bibfnamefont {Alejandro}\ \bibnamefont {Boh\'e}},\
  }\bibfield  {title} {\enquote {\bibinfo {title} {{Frequency-domain
  gravitational waves from nonprecessing black-hole binaries. II. A
  phenomenological model for the advanced detector era}},}\ }\href {\doibase
  10.1103/PhysRevD.93.044007} {\bibfield  {journal} {\bibinfo  {journal} {Phys.
  Rev. D}\ }\textbf {\bibinfo {volume} {93}},\ \bibinfo {pages} {044007}
  (\bibinfo {year} {2016})},\ \Eprint {http://arxiv.org/abs/1508.07253}
  {arXiv:1508.07253 [gr-qc]} \BibitemShut {NoStop}%
\bibitem [{\citenamefont {Dhurkunde}\ \emph {et~al.}(2022)\citenamefont
  {Dhurkunde}, \citenamefont {Fehrmann},\ and\ \citenamefont
  {Nitz}}]{Dhurkunde:2021csz}%
  \BibitemOpen
  \bibfield  {author} {\bibinfo {author} {\bibfnamefont {Rahul}\ \bibnamefont
  {Dhurkunde}}, \bibinfo {author} {\bibfnamefont {Henning}\ \bibnamefont
  {Fehrmann}}, \ and\ \bibinfo {author} {\bibfnamefont {Alexander~H.}\
  \bibnamefont {Nitz}},\ }\bibfield  {title} {\enquote {\bibinfo {title}
  {{Hierarchical approach to matched filtering using a reduced basis}},}\
  }\href {\doibase 10.1103/PhysRevD.105.103001} {\bibfield  {journal} {\bibinfo
   {journal} {Phys. Rev. D}\ }\textbf {\bibinfo {volume} {105}},\ \bibinfo
  {pages} {103001} (\bibinfo {year} {2022})},\ \Eprint
  {http://arxiv.org/abs/2110.13115} {arXiv:2110.13115 [astro-ph.IM]}
  \BibitemShut {NoStop}%
\bibitem [{\citenamefont {Soni}\ \emph {et~al.}(2022)\citenamefont {Soni},
  \citenamefont {Gadre}, \citenamefont {Mitra},\ and\ \citenamefont
  {Dhurandhar}}]{Soni:2021vls}%
  \BibitemOpen
  \bibfield  {author} {\bibinfo {author} {\bibfnamefont {Kanchan}\ \bibnamefont
  {Soni}}, \bibinfo {author} {\bibfnamefont {Bhooshan~Uday}\ \bibnamefont
  {Gadre}}, \bibinfo {author} {\bibfnamefont {Sanjit}\ \bibnamefont {Mitra}}, \
  and\ \bibinfo {author} {\bibfnamefont {Sanjeev}\ \bibnamefont {Dhurandhar}},\
  }\bibfield  {title} {\enquote {\bibinfo {title} {{Hierarchical search for
  compact binary coalescences in the Advanced LIGO\textquoteright{}s first two
  observing runs}},}\ }\href {\doibase 10.1103/PhysRevD.105.064005} {\bibfield
  {journal} {\bibinfo  {journal} {Phys. Rev. D}\ }\textbf {\bibinfo {volume}
  {105}},\ \bibinfo {pages} {064005} (\bibinfo {year} {2022})},\ \Eprint
  {http://arxiv.org/abs/2106.08925} {arXiv:2106.08925 [gr-qc]} \BibitemShut
  {NoStop}%
\end{thebibliography}%

\end{document}